\documentstyle[epsfig]{mn}

\input{epsf}

\def\lsim{\mathrel{\hbox{\rlap{\hbox{\lower4pt\hbox{$\sim$}}}\hbox{$<$}}}}

\def\gsim{\mathrel{\hbox{\rlap{\hbox{\lower4pt\hbox{$\sim$}}}\hbox{$>$}}}}

\def\and   {\rm {et al.} \rm}  

\begin{document}

\title
[The QSO Power Spectrum from the 10k Catalogue]
{
The 2dF QSO Redshift Survey - IV. The QSO Power Spectrum from the 10k Catalogue
}

\author[F. Hoyle  et al ]
{
Fiona Hoyle$^{1,*}$, P.~J. Outram$^1$, T. Shanks$^1$,  S. M. Croom$^{2}$, B.~J. Boyle$^{2}$ 
\newauthor   N.~S. Loaring$^{3}$, L. Miller$^{3}$ \& R.~J. Smith$^{4}$,
\\
1 Department of Physics, Science Laboratories, South Road, Durham, DH1 3LE, U.K
\\
2 Anglo-Australian Observatory, PO Box 296, Epping, NSW 2121, Australia\\
3 Department of Physics, Oxford University, Keble Road, Oxford, OX1 3RH, UK\\
4 Liverpool John Moores University, Twelve Quays House, Egerton Wharf, Birkenhead, CH41 1LD \\
* email hoyle@venus.physics.drexel.edu\\
}

\maketitle 
\vspace{-1cm}
 
\begin{abstract}

We present a power spectrum analysis of the 10K catalogue from 
the 2dF QSO Redshift Survey. Although the Survey currently has a patchy
angular selection function, we use the Virgo Consortium's Hubble Volume
simulation to demonstrate that we are able to make a useful first
measurement of the power spectrum over a wide range of scales. 

We compare the redshift-space power spectra of QSOs to those measured for
galaxies and Abell clusters at low redshift and find that they show similar
shapes in their overlap range, 50-150h$^{-1}$Mpc, with
P$_{\rm QSO}(k)\propto k^{-1.3}$. The amplitude of the QSO power spectrum at
$z \approx$ 1.4 is almost comparable to that of galaxies at the present day if
$\Omega_{\rm m}$=0.3 and $\Omega_\Lambda$=0.7 (the $\Lambda$ cosmology), 
and a factor of $\approx$ 3 lower
if $\Omega_{\rm m}$=1 (the EdS cosmology) is assumed. 
The amplitude of the QSO power spectrum is a
factor of $\approx$ 10 lower than that measured for Abell clusters at the
present day. At larger scales, the QSO power spectra continue to rise robustly
to $\approx$ 400 h$^{-1}$Mpc, implying more power at large
scales than in the APM galaxy power spectrum measured by Baugh \&
Efstathiou. 

We split the QSO sample into two redshift bins and find little evolution
in the amplitude of the power spectrum, consistent with the result for the
QSO correlation function. In models with $\Omega_{\rm m} \gsim $0.1 
this represents
evidence for a QSO-mass bias that evolves as a function of time. 

We compare the QSO power spectra to CDM models to obtain a
constraint on the shape parameter, $\Gamma$. For two choices of
cosmology ($\Omega_{\rm m}$=1, $\Omega_\Lambda$=0 and 
$\Omega_{\rm m}$=0.3, $\Omega_\Lambda$=0.7), we 
find the best fit model has $\Gamma \approx 0.1 \pm0.1$.
In addition, we have shown that a power spectrum  analysis of the
Hubble Volume $\Lambda$CDM mock QSO catalogues with $\Gamma=0.17$
as input, produces a result which is statistically consistent with
the data. The analysis of the mock catalogues also indicates that the
above results for $\Gamma$ are unlikely to be dominated by systematic 
effects due to the current catalogue window. 
We conclude that the form of the QSO power spectrum shows
large-scale power significantly in excess of the standard CDM prediction,
similar to that seen in local galaxy surveys at intermediate scales.

\end{abstract}


\section{Introduction}

Galaxy surveys have allowed us to study the large scale structure in
the local Universe. Large clusters of galaxies and voids have been
detected in these surveys and they have placed constraints on
cosmological parameters and models of structure formation. However,
large area ($> 100$ deg$^2$), 
galaxy redshift surveys have so far only probed clustering
out to $z\lsim 0.1$. Even the 2dF Galaxy Redshift Survey
(Colless 1998) and the Sloan Digital Sky Survey (Gunn \& Weinberg 1999) will
only observe galaxies with redshifts $\lsim 0.3$. 
Surveys such as the Canada-France Redshift Survey (Crampton et al. 1995) 
and the
CNOC Survey (Yee et al. 2000) have probed galaxy clustering out past $z=0.5$
and, in the future, the DEEP (Davis \& Faber 1998) and VIRMOS 
(Le F\`{e}vre et al. 1998)
Surveys will probe galaxy clustering out past $z \approx 1$. Steidel and
collaborators have even developed techniques from which galaxies can
be selected at higher redshifts by looking for the Lyman break 
(Steidel et al. 1995, 1999). However these deep surveys currently only
cover a small area of sky and the resulting `pencil beams' are
unsuitable for direct power spectrum measurements. If we wish to
measure the power spectrum of clustering out to high redshifts over a
wide area in a reasonably
short time period, then different techniques or a different class of
object are required.

Since the work of Osmer (1981), redshift surveys of
Quasi-Stellar Objects (QSOs) have been used to probe clustering at
high redshifts. The initial attempts to measure QSO clustering proved
inconclusive, due to the low number density of QSOs and the
inhomogeneous nature of the surveys. However, it was demonstrated in
the mid-1980's that QSOs are clustered, at least on small $r<10
h^{-1}$Mpc scales, and that the amplitude of QSO clustering is similar
to that of present day, optically selected galaxies 
(Shaver 1984, Shanks et al. 1986, 1987).
More recently, several groups have
looked at the evolution of QSO clustering. Croom \& Shanks (1996) found 
that the amplitude of QSO clustering remains approximately constant with 
redshift, whereas La Franca, Andreani \& Cristiani (1998) found that
the clustering amplitude of optically bright QSOs evolves slowly with 
redshift, such that the clustering amplitude of high redshift QSOs is 
2$\sigma$ higher than that of low redshift QSOs. However, existing
surveys of QSOs are still fairly small, for example, the Durham/AAT Survey 
(Boyle 1986) contains $\approx 400$ QSOs and 
Large Bright QSO Survey (Hewitt, Foltz \& Chaffee 1995) contains 1,053 QSOs.
Measurements of some clustering statistics have been hampered by the 
small number of objects contained in the survey or the limited sky 
coverage of the surveys.

Our knowledge of the structure of the Universe at high redshifts will
be drastically improved once the 2dF QSO Redshift Survey 
(2QZ, Croom et al. 1998; Boyle et al. 2000; Croom et al. 2001)
and the QSO part of the 
Sloan Digital Sky Survey (Gunn \& Weinberg 1995) are finished.
 These surveys will 
contain at least a factor of 25 more QSOs and cover a larger area of
sky than existing QSO surveys. One of the key aims of both surveys is
to measure the clustering of QSOs out to $\approx 1000 h^{-1}$Mpc with
greater accuracy than is possible with existing surveys. For
example, the increased number of QSOs in the 2dF QSO survey should 
reduce the errors on the measured correlation function by at least a factor 
of 5 on scales $\gsim 10 h^{-1}$Mpc (Croom et al. 2001)

The power spectrum is now established as one of the favoured methods
of quantifying clustering. 
(Baugh \& Efstathiou 1994; Tadros \& Efstathiou 1996; Lin et al. 1996;
Hoyle et al. 1999).
One advantage of the power spectrum over the two-point correlation 
function is that the two-point correlation function is affected by 
uncertainties in the mean density on all scales, whereas the power 
spectrum is only affected by these uncertainties on the largest scales 
(Cole, Fisher \& Weinberg 1995). 
Ideally, we would like to measure the power spectrum of 
the mass density field, as this is the statistic predicted by models 
of structure formation. However, we are only able to observe the light 
in the Universe. Therefore, we generally measure the power spectrum of objects
such as galaxies, clusters and QSOs, although there are promising 
methods for measuring the mass power spectrum from Lyman alpha forest
systems (Croft et al. 1998) and from weak lensing (Wilson, Kaiser 
\& Luppino 2001). 

However, if the bias is scale independent, then although the amplitude
of the QSO power spectrum may be different to that of the mass, the
form of the QSO power spectrum will directly probe the shape of the
mass power spectrum; the assumption of scale independent bias 
may be more appropriate at large scales (Coles 1993; 
Mann, Peacock \& Heavens 1998).

The power spectrum has never before been measured from optically
selected QSOs. QSOs are detected out to high redshifts and so QSO
surveys have a large volume. However, existing surveys only cover 
small areas of sky and contain only a small number of QSOs. The survey 
geometry and low QSO number density within the surveys would have 
severely restricted the range of scales over which the power spectrum 
could have been reliably measured. 

In this paper, we apply a power spectrum analysis to the 10k release
catalogue.
A correlation function analysis of the 
identical sample has been performed by Croom et al. (2000). We describe 
the current data set in Section 
\ref{sec:qsodata} and look at the angular distribution of the QSOs 
in order to create the random catalogues needed in the power spectrum 
analysis. In Section \ref{sec:pkqso} we briefly describe the method 
of power spectrum analysis and in Section \ref{sec:test} we test the
measurement of the power spectrum from the current survey window using
mock catalogues drawn from the Virgo consortiums {\it Hubble Volume}
lightcone simulation (Frenk et al. 2000). In Section \ref{sec:qsoreslts}
we compare the QSO power spectrum to power spectra of present day 
galaxies and clusters and we see how the QSO power spectrum evolves 
with redshift. In Section \ref{sec:qsoimp} we compare the QSO power 
spectrum with power spectra of models of large scale structure and 
in Section \ref{sec:qsoconc} we draw our conclusions.

\section{The 10k Catalogue}
\label{sec:qsodata}

In this paper, we analyse QSOs that will be contained in the first public 
release of the 2QZ, known as the {\it 10k Catalogue}. The catalogue will 
contain 10681 QSOs and will be released to the community in the 
first half of 2001. This catalogue will contain the most spectroscopically 
complete fields (i.e. fields in which more than 85 per cent of objects have 
been identified) that were observed prior to November 2000. It
will eventually be available from {\tt www.2dfquasar.org}.

QSO candidates are identified from broad band 
{\it ub$_{\rm J}$r} colours from Automatic Plate Measuring
(APM) facility measurements of UK Schmidt Telescope (UKST) photographic 
plates. This colour selection gives a photometric completeness of
better that 90 per cent up to $z \approx 2.2$ (Boyle et al. 2000).
The spectra of the objects are obtained using the 2dF 
instrument on the AAT. The spectra are reduced using the 2dF pipeline reduction
system (Bailey \& Glazebrook 1999) and objects are
identified as QSOs by an automated procedure known as {\tt AUTOZ} 
(Miller et al. 2001). {\tt AUTOZ} also identifies the QSO redshifts; these 
have also been visually checked by two independent observers.

Observations have been made over two 5$\times$75 deg$^2$ strips in the
North and South Galactic Cap regions. The North
Galactic Cap strip is centred at $\delta= 0^{\circ}$ with
09$^{\rm h}$50$^{\rm m}$ $\lsim \alpha \lsim $ 14$^{\rm h}$50$^{\rm m}$, 
while the South Galactic Cap strip
is centred at $\delta = -30^{\circ}$, with 21$^{\rm h}$40$^{\rm m}$ 
$\lsim \alpha \lsim $ 03$^{\rm h}$15$^{\rm m}$. We 
will refer to these regions as the NGC and SGC respectively. The finished survey
will cover an area of 740 deg$^2$. See Croom (1997), Smith (1998) and Smith et al. (2001) for further details of the photometric catalogue.

For each QSO, we have an angular position and a redshift. In order to
convert from a redshift into a comoving distance, we need to adopt a
cosmology. We consider an $\Omega_{\rm m}$=1.0, $\Omega_{\Lambda}$=0.0 
cosmology (EdS hereafter) and an
$\Omega_{\rm m}$=0.3, $\Omega_{\Lambda}$=0.7 cosmology (referred to as
the $\Lambda$ cosmology hereafter). Other cosmologies, such as open
cosmologies, could also be considered but here we restrict our
analysis to flat cosmologies, consistent with the recent results from
balloon experiments such as Boomerang (de Bernardis et al. 2000) and Maxima
(Balbi et al. 2000).

\subsection{Constructing the Random Catalogue}
\label{sec:rancat}
\subsubsection{The Radial Selection Function}

\begin{figure} 
\begin{centering}
\begin{tabular}{c}
{\epsfxsize=8truecm \epsfysize=8truecm \epsfbox[35 170 550 675]{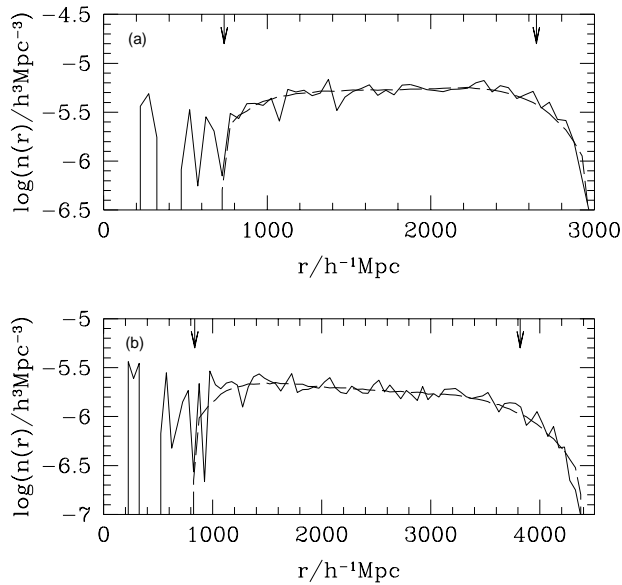}}
\end{tabular}
\caption
{ The radial number density of QSOs in bins of $\Delta r = 50
h^{-1}$Mpc in the SGC (solid line) calculated assuming EdS (a) and
$\Lambda$ (b). The dashed lines show the number density as measured
from the random catalogues, normalised to match the number density of
the QSOs. The number density varies slowly as a function of scale,
decreasing only by a factor of approximately three over the range $0.3<z<2.2$, 
indicated by the arrows. At higher redshifts, the number density
decreases rapidly.}
\label{fig:nr}
\end{centering}
\end{figure}

The power spectrum measured directly from a QSO redshift
survey is a convolution of the power spectrum of QSO clustering with
the power spectrum of the survey geometry. In order to estimate the
power spectrum of the survey geometry, or window function, we generate
a catalogue of random points which have the same angular and 
radial selection function as the QSOs but otherwise are unclustered. 

Redshifts are drawn at random according to a 6th order polynomial $N(z)$ fit
over the range of redshift and geometry of the sample under consideration.
Using an approximation to the true $N(z)$ distribution can remove some
of the large scale power. We test this by creating a 10th order
polynomial $N(z)$ fit and remeasure the $\Lambda$ cosmology power spectrum. 
Out to scales of 200$h^{-1}$Mpc (log$(k / h \,{\rm Mpc}^{-1} ) > -1.5 $) the two power spectra
are virtually indistinguishable. At 400$h^{-1}$Mpc 
(log$(k / h \,{\rm Mpc}^{-1} ) > -1.8 $), the
power spectra agree in amplitude to within 15\%. The fractional error at this
scale is 40\% so the difference is negligible compared to the errors.
The Survey geometry is also limiting how accurately the power spectrum can be
measured on these scales too. (See 
Section \ref{sec:test} for details of the window function tests.)

Figure \ref{fig:nr} shows the radial selection function of QSOs in the
SGC (solid line) and in the random catalogue (dashed line) 
assuming EdS (a) and assuming
the $\Lambda$ cosmology (b). Over a wide range of redshift
the selection function of the random catalogue accurately 
reproduces that
of the QSOs. The radial selection function of the QSOs
only varies slowly as a function of scale 
over the range of redshifts $0.3<z<2.2$, indicated by
the arrows. The value of $n(r)$ changes by only a factor of
$\approx$3 over the range of scales corresponding to $0.3<z<2.2$,
whereas the value of $n(r)$ for galaxies 
changes by a far greater factor over a far smaller range of
scales (for example see Figure 2 in Hoyle et al. 1999). 
This is because the luminosity evolution
approximately cancels the effect of luminosity distance so we are
always surveying the luminosity function to approximately the same
absolute magnitude, relative to the break, $M_B^*$, 
almost independent of redshift. 
As it is desirable to have a constant number density when 
measuring clustering statistics to remove the need for a weighting scheme
(other than giving all QSOs equal weight),
we restrict the QSOs in our sample to have redshifts in the range 
$0.3 < z < 2.2$. This gives us a sample of 8935 QSOs.

\subsubsection{The Angular Selection Function}
\label{sec:rancat_ang}

Observations have been made across both the NGC and SGC strips. An
optimal tiling algorithm for the 2dF Galaxy Redshift survey was
developed by the 2dF Consortium to allow as many galaxies and QSOs as
possible to be observed in each pointing. Pointings also overlap in high
density regions to maximise the coverage in all areas of the survey.
Therefore, the 10k catalogue has a patchy angular selection function, 
as can be seen in Figure 1. of Croom et al. (2001), which has to be
matched by the random catalogue.

To match the angular selection function of the QSOs, we construct a
completeness map. 
In each region defined by the intersection of 2dF pointings we calculate the
number of objects observed and
divide this by the number of candidates within that region to define the
fractional observational completeness. This allows us to estimate the   
completeness in areas where two or more 2dF fields overlap. The
completeness map is then rebinned to 1'x1' bins and the number of random 
points in each bin is weighted by this fractional completeness.

We also account for the extinction due to
galactic dust as this changes the effective magnitude limit of the
survey as a function of position. We use the estimate of the dust
reddening, E(B-V), as a function of position given by 
Schlegel, Finkbeiner \& Davis (1998) and
weight the random distribution by
\begin{equation}
W_{\rm ext}(\alpha,\delta) = 10^{- \beta {\rm A_{\rm b_J}}}
\end{equation}
where ${\rm A_{b_j}}$ = 4.035E(B-V) and $\beta=0.3$, the slope of the
QSO number counts at the magnitude limit of the survey, b$_{\rm
j}$=20.85. We have measured the power spectrum using random catalogues
that were constructed with or without taking into account the effect of dust. 
We find that dust has a less than 5\% effect on the
power spectrum amplitude out to scales of 300$h^{-1}$Mpc 
(log$(k / h \,{\rm Mpc}^{-1} ) > -1.7 $) for the EdS cosmology, 
400$h^{-1}$Mpc (log$(k / h \,{\rm Mpc}^{-1} ) > -1.8 $) for the 
$\Lambda$ cosmology.

One further effect that could influence the shape of the power
spectrum on large scales is due to possible errors in the zero-point 
calibration of the UKST plates. A conservative estimate of the current 
uncertainty is $\approx$0.2 mag (Smith et al. 2001), although further CCD 
photometry is currently being obtained and we expect that the 
uncertainty in the zero-point calibration will be considerably 
lower in the final catalogue. 

With a QSO number count slope of $\beta$=0.3, a 0.2 mag change in the
zero-point calibration would change the expected QSO number density 
by $\approx$15 per cent. 
To estimate the effect this has on the QSO power spectrum, we 
generate different realisations of the random catalogues 
including plate-to-plate
variations in the zero-point calibration. These are introduced by
randomly varying the number density of points in the random catalogue on
each UKST plate by r.m.s. 15 per cent. The uncertainty this introduces
into the QSO power spectrum estimate is then given by the dispersion in
the power spectra obtained from each realisation. We find that the
fractional error introduced by this uncertainty varies between 1 and 5 per
cent up to scales of 300$h^{-1}$Mpc or log$(k / h \,{\rm Mpc}^{-1} ) > -1.7 $
(400$h^{-1}$Mpc or log$(k / h \,{\rm Mpc}^{-1} ) > -1.8 $ in 
the $\Lambda$ case), 
where the QSO power spectrum can be robustly measured. On these scales the
fractional FKP errors (discussed below) are $\approx$30 per cent, 
and therefore the total error
is increased by $\lsim 1\%$, less than the potential systematic error.

\begin{figure} 
\begin{centering}
\begin{tabular}{c}
{\epsfxsize=8truecm \epsfysize=8truecm \epsfbox[35 170 550 675]{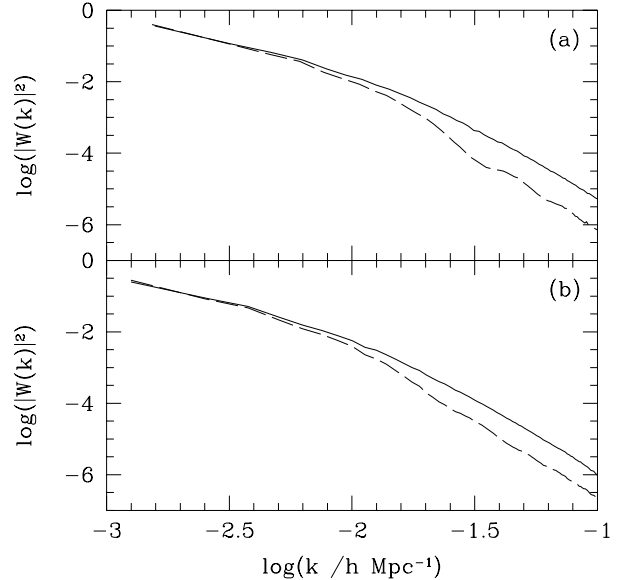}}
\end{tabular}
\caption
{The window function of the current 2QZ in the SGC (solid line)
compared to the {\it expected} window function of a finished
5$\times 75$ deg$^2$ strip (dashed line) calculated assuming
EdS (a) and $\Lambda$ (b). The window function of the NGC is very
similar to that of the SGC and both are a steep power law out to
scales of $\approx 300 h^{-1}$Mpc ($\approx 400 h^{-1}$Mpc for the $\Lambda$
cosmology) as opposed to $\approx 450 h^{-1}$Mpc (600$h^{-1}$Mpc) when the
survey is finished.}

\label{fig:qsowf}
\end{centering}
\end{figure}

In Figure \ref{fig:qsowf} we show
the window function of the current 2QZ in the SGC (solid line)
compared to the {\it expected} window function of a finished
5$\times 75$ deg$^2$ strip (dashed line) calculated assuming EdS (a) and
$\Lambda$ (b). The window function of the NGC is very similar in slope and 
amplitude as although more QSOs have been observed on the SGC strip, the QSOs
in the NGC are evenly distributed across the $5 \times 75$ deg$^2$ strip
(see Figure 1 Croom et. al 2001b). Both window
functions are steep power laws, proportional to $k^{-4}$ out to scales of 
$\approx 300h^{-1}$Mpc (log$(k / h \,{\rm Mpc}^{-1} ) > -1.7 $) EdS 
or $\approx 400 h^{-1}$Mpc 
(log$(k / h \,{\rm Mpc}^{-1} ) > -1.8 $) for the $\Lambda$ cosmology.
When the survey is finished, the power laws will extend to
$\approx 500 h^{-1}$Mpc (log$(k / h \,{\rm Mpc}^{-1} ) > -1.9 $) EdS,
600$h^{-1}$Mpc (log$(k / h \,{\rm Mpc}^{-1} ) > -1.98 $) for the 
$\Lambda$ cosmology (see Hoyle 2000).
We conclude that geometric effects (dust, plate-to-plate variations in the 
zero points, window function of the survey) do not effect the shape of the QSO
power spectrum on scales $\lsim 300 h^{-1}$Mpc 
(log$(k / h \,{\rm Mpc}^{-1} ) > -1.7 $) assuming the EdS cosmology 
$\lsim 400 h^{-1}$Mpc (log$(k / h \,{\rm Mpc}^{-1} ) > -1.8 $ )
for the $\Lambda$ cosmology.

\section{Measuring the QSO Power Spectrum}
\label{sec:pkqso}

\subsection{Power Spectrum Estimation}

The power spectrum estimator that we adopt is similar to
that given Tadros \& Efstathiou (1996) and the same as that given in
Hoyle et al. (1999). We outline the details below.

The Fourier transform of the observed QSO density field, 
within a periodic volume $V$, is given by 
\begin{equation}
\hat{n}_{\circ}({\bf k}) = \frac{1}{V} \sum_i {\rm e}^{\rm i \textbf{k.x$_i$(r)}}.
\end{equation}
where the ${\bf x_i(r)}$ refer to the spatial positions of the QSOs. 
The Fourier transform of the survey
window function is approximated by:
\begin{equation}
\hat{W}_e({\bf k}) = \frac{1}{V} \sum_i {\rm e}^{\rm i\textbf{k.x$_i$(r)}},
\end{equation}
where this time the ${\bf x_i(r)}$ refer to the spatial positions of the random 
points
(cf equations 7, 8 and 9 of Tadros \& Efstathiou 1996 and equations 3 and
4 in Hoyle et al. 1999).
Due to the large number of unclustered random points used, shot noise 
makes a negligible contribution to this estimate. 
The power spectra of the survey window function, shown in
Figure \ref{fig:qsowf}, are much steeper than the expected QSO
power spectrum, falling off as $\propto k^{-4}$ for wavenumbers
log$(k / h \,{\rm Mpc}^{-1} ) > -1.7 $ or scales of 300$h^{-1}$Mpc in the 
EdS case (log$(k / h \,{\rm Mpc}^{-1} ) > -1.8 $ or scales of 400$h^{-1}$Mpc 
in the $\Lambda$ case).
Therefore the main effect of the convolution with the
survey window function is to alter the shape of the power spectrum only
on scales larger than 300$h^{-1}$Mpc, log$(k / h \,{\rm Mpc}^{-1} ) > -1.7 $, (400 $h^{-1}$Mpc, log$(k / h \,{\rm Mpc}^{-1} ) > -1.8 $ assuming the
$\Lambda$ cosmology).

Following Tadros \& Efstathiou, we define a quantity with a 
mean value of zero:
\begin{equation}
\delta({\bf k})  = \hat{n}_{\circ}({\bf k}) - \alpha \hat{W}_{e}({\bf k}),
\label{eq:pkdef}
\end{equation}
where $\alpha$ is the ratio of the number of QSOs to random points
in samples with constant number density. The power spectrum of QSO clustering is
then estimated using:
 
\begin{eqnarray}
P_{e} (k) &=& \left[  \frac{1}{ N_{\rm ran}^{2} }\frac{1}{V} 
 \sum_{k'} \left( V^2\left| W_{e}(k') \right|^{2} - 
\frac{1}{ N_{\rm ran} } \right) \right]^{-1} 
\nonumber \\
&\times& \left( \frac{ V^{2} |\delta(k)|^{2} }{ N_{\rm QSO}^{2} } 
- \frac{1}{ N_{\rm QSO} } - \frac{1}{ N_{\rm ran} } \right) .
\label{eq:pest}
\end{eqnarray}
We use a Fast Fourier Transform (FFT)
to compute the power spectrum and bin the Fourier
modes up logarithmically to reduce the covariance between each bin.

The limit beyond which a power spectrum cannot be measured using a FFT is
known as the Nyquist frequency. This is the frequency at which the grid  
samples the wave exactly twice per cycle:
\begin{equation}
k_{\rm Nyq} = \frac{1}{2} \frac{2\pi}{\Delta},
\end{equation}
where, for a 256$^3$ FFT and a box of size 5000$h^{-1}$Mpc, the interval
$\Delta$=5000/256 $h^{-1}$Mpc. The limit out to which the power spectrum  
measurement is reliable depends on the grid assignment scheme, and is a
fraction of k$_{\rm Nyq}$. Using the nearest grid point assignment
scheme, $k_{\rm lim} \approx 1/2$ $k_{\rm Nyq}$ (Hatton 1999), 
which, for the above box size, corresponds to
 a value of log$(k_{\rm lim}) \approx  -1$ or a scale of $\approx 60 h^{-1}$Mpc.

To enable measurements of
the power spectrum on smaller scales, we use a direct method of
computing the Fourier transform.
This method is time
intensive and can only be applied to a small number of particles (so
cannot be applied to the random catalogue). For the direct method, we make
the assumption that equation \ref{eq:pkdef} can be rewritten as
$\delta({\bf k}) = \hat{n}({\bf k})$,
which is only
valid on small scales where the window function has a small amplitude
(Figure \ref{fig:qsowf}). Using mock catalogues drawn from the {\it
Hubble Volume} simulation (see Section \ref{sec:test}), we
find that on scales smaller than $\approx 100 h^{-1}$Mpc (log$(k) \approx
-1.2$) this approximation is accurate.

\subsection{Error Determination}
\label{sec:errors}

The errors on the power spectrum are estimated using the method of 
Feldman, Kaiser \& Peacock (1994), 
equation 2.3.2, assuming that all QSOs carry equal 
weight when
estimating the QSO power spectrum. The error is given by
\begin{equation} 
\frac{\sigma^2(k)}{P^2(k)} = \frac{(2\pi)^3 [1 +\frac{1}{n(r)P(k)}]^2}{V_k V_s}
\label{eq:fkper}
\end{equation}
where $V_k$ is the volume of each bin in $k$-space, estimated by
$V_k=  N_k (\Delta k)^3 $ with $N_k$ the number of independent modes in the
$k$-shell and $(\Delta k)^3$ the volume of one $k$-mode. $V_s$ is the
volume of the survey. Over a wide range of scales these errors are
found to be in reasonable agreement with errors estimated from 
the dispersion over power spectrum estimates from the {\it Hubble Volume} mock
catalogues (see Section \ref{sec:test} and Hoyle 2000). On the
smallest scales, the FKP errors appear smaller
than those from the mock catalogues. 
FKP errors are only valid in the fully linear regime and non-linearities 
on small scales may mean they  underestimate the true error on these 
scales Meiksin \& White (1999).

To obtain a single 2QZ power spectrum for each assumed cosmology, we
average the power spectrum of the QSOs in the NGC and SGC together, weighting by the inverse of the variance on each scale.

\begin{figure} 
\begin{centering}
\begin{tabular}{c}
{\epsfxsize=8truecm \epsfysize=8truecm \epsfbox[35 170 550 675]{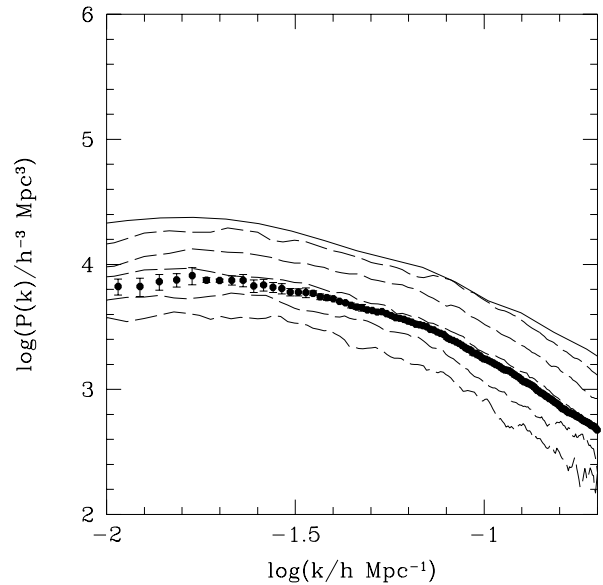}} \\
\end{tabular}
\caption{The solid line shows the input power spectrum to the {\it
Hubble Volume} simulation. The dashed lines show power spectra of the
mass in the simulation in bins of 0.5 in redshift, centred on
(reading down) 0.25, 0.75, 1.25, 1.75 and 2.25. The points show the power
spectrum of the mass in the light cone out to $z<2.4$. The
amplitude of this power spectrum is close to the amplitude of the
power spectrum at the average redshift of the sample which is $\bar{z}
= 1.4$.}
\label{fig:light}
\end{centering}
\end{figure}

\begin{figure} 
\begin{centering}
\begin{tabular}{c}
{\epsfxsize=8truecm \epsfysize=8truecm \epsfbox[35 170 550 675]{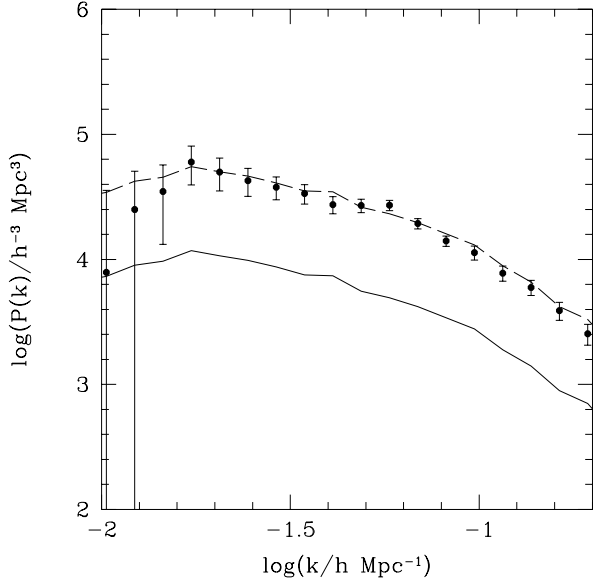}} \\
\end{tabular}
\caption{
The points show the power
spectrum measured from the biased mock catalogues, the solid line shows the
mass power spectrum and the dashed line shows the mass power spectrum
multiplied by a factor of $b^2$=4.7. Both the mass and the biased
catalogues have the selection function of the QSOs imprinted on them
and are measured in redshift space.}
\label{fig:pkbias}
\end{centering}
\end{figure}

\section{Testing P(k) estimators using the Hubble Volume}
\label{sec:test}

To test the estimator of the power spectrum and the effect of the
window function on the recovered power spectrum, we have constructed
mock catalogues that approximately match the clustering and geometry 
expected from the final QSO sample. Full details of the mock
catalogues can be found in Hoyle (2000). We review the
features most pertinent to the analysis here.

The mock catalogues are constructed from the Virgo consortium's {\it
Hubble Volume} simulation (see Frenk et al. 2000). 
The simulation has a $\Lambda$CDM cosmology with 
$\Omega_{\rm m}$=0.3, $\Omega_b$=0.04 and $\Omega_{\Lambda}$=0.7, $h$=0.7
and normalised to $\sigma_8$=0.9 at
present day and with shape parameter of $\Gamma = 0.17$. 
The simulation has a lightcone 
output such that the evolution of the dark matter is fully accounted for. It
extends to redshift $\approx 4$, 
and covers an area of 15$\times$75deg$^2$, which we split 
into three 5$\times$75deg$^2$ strips to match the geometry of the 2QZ strips.
The three strips are not entirely independent so when we imprint the
NGC and SGC completeness maps (see later in this section) on the simulations,
we just consider the outer slices.

The simulation has a lightcone
output such that the evolution of the dark matter is fully accounted
for. However, in Figure \ref{fig:light} we show that evolution has
little effect on the shape of the mass power spectrum on the scales where 
it can be robustly measured. The
solid line is the input power spectrum to the {\it Hubble Volume}
simulation. The dashed lines show the power spectra of the mass in the
simulation in bins of 0.5 in redshift centred on (reading down) 0.25,
0.75, 1.25, 1.75 and 2.25. The points show the power spectrum of the
mass over the light cone out to $z < 2.5$, the average redshift of
the mass is $z \approx 1.4$. This power spectrum does indeed lie between the power
spectra centred on $z=1.25$ and $z=1.75$ and all three power spectra
have very similar shapes.

We bias the dark matter particles to approximately match the
clustering expected from 2QZ. Initial measurements of the
correlation function suggested that the correlation length of the 2QZ
QSOs would be $\approx 6h^{-1}$Mpc, assuming the $\Lambda$ cosmology,
and that the comoving correlation length would be roughly constant with redshift. See Croom et al. (2001) for the latest measurements of the correlation
function.

\begin{figure*} 
\begin{centering}
\begin{tabular}{ccc}
{\epsfxsize=5.5truecm \epsfysize=5.5truecm \epsfbox[35 170 550
675]{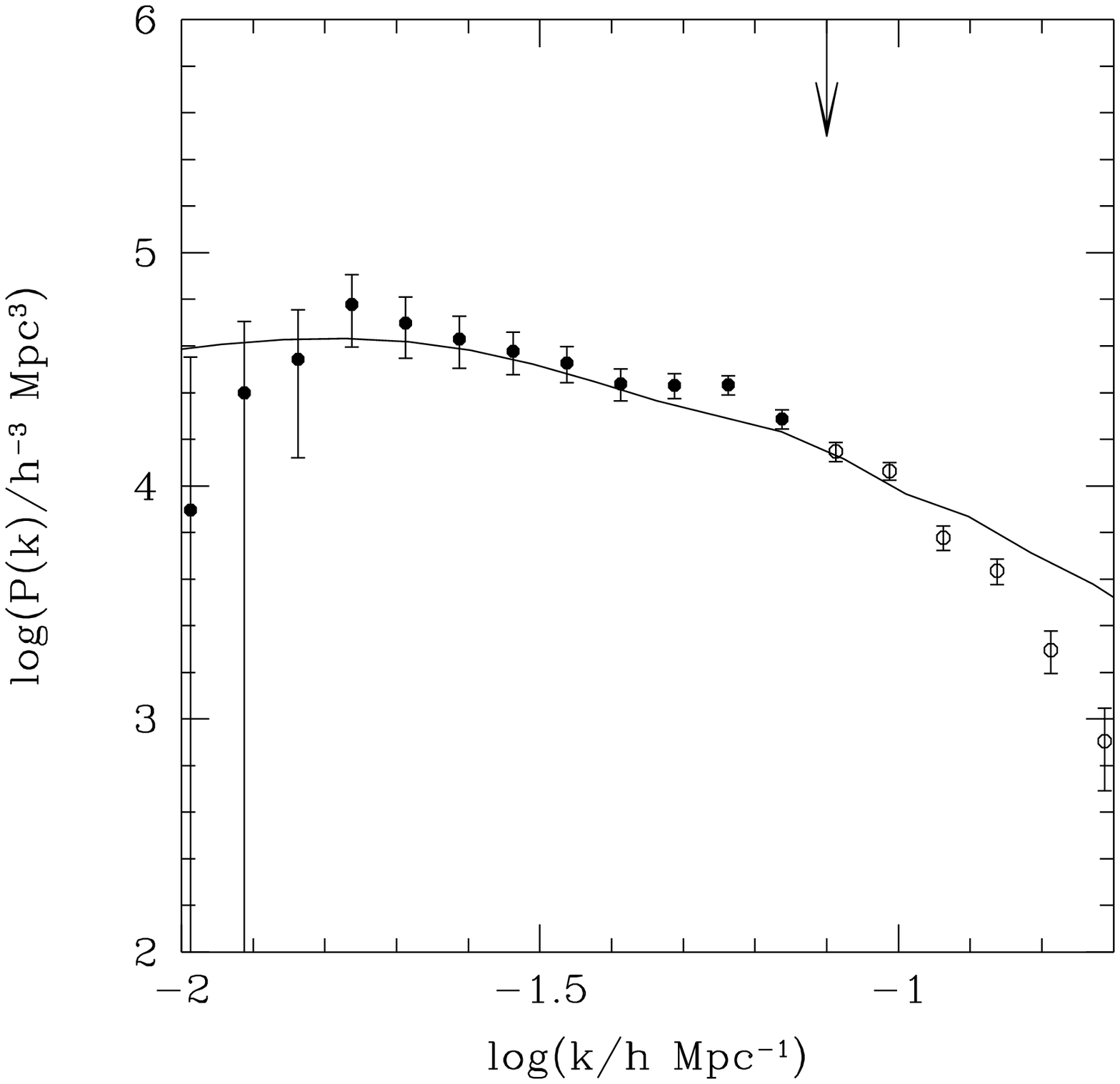}} &
{\epsfxsize=5.5truecm \epsfysize=5.5truecm \epsfbox[35 170 550
675]{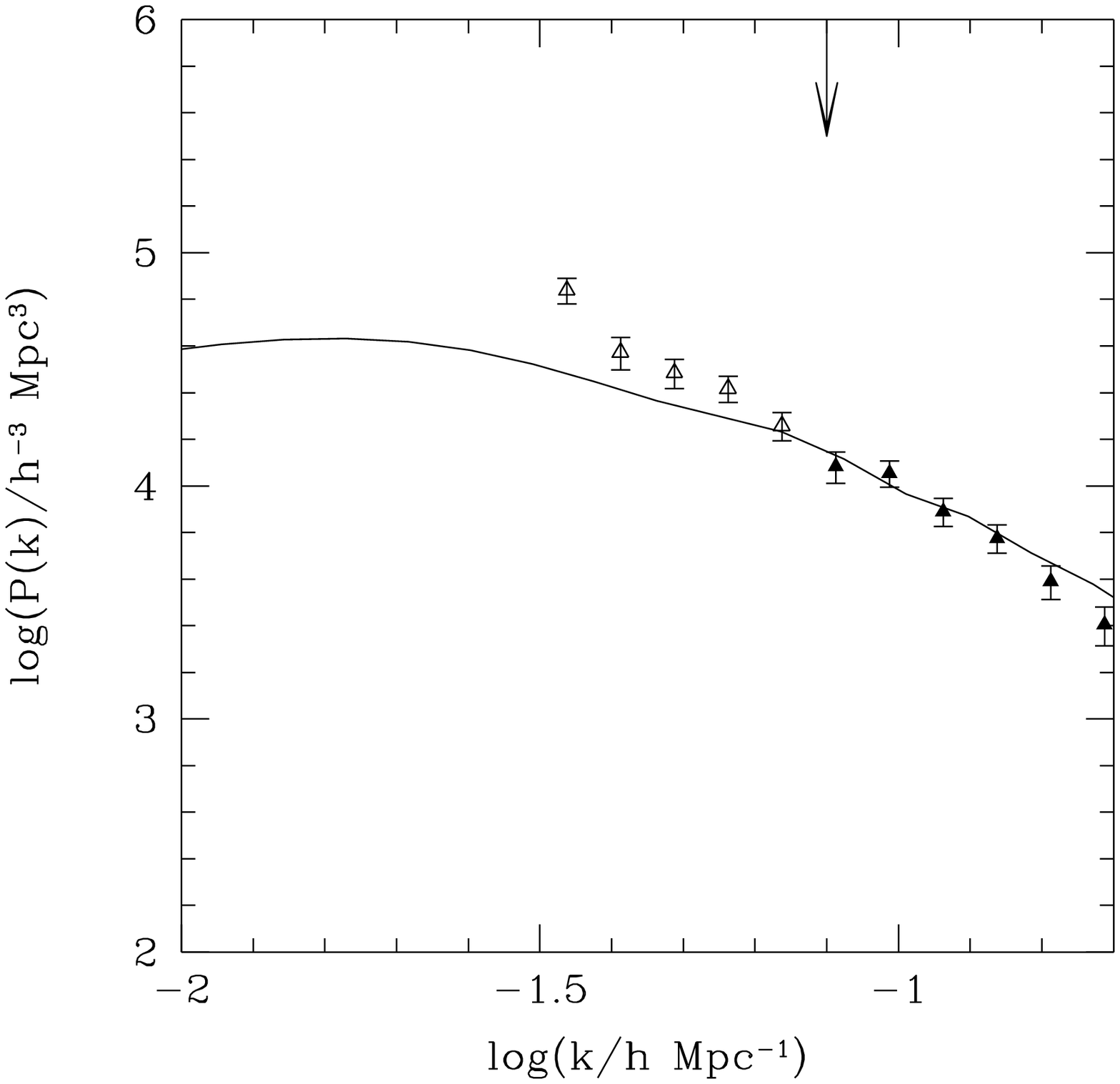}} &
{\epsfxsize=5.5truecm \epsfysize=5.5truecm \epsfbox[35 170 550
675]{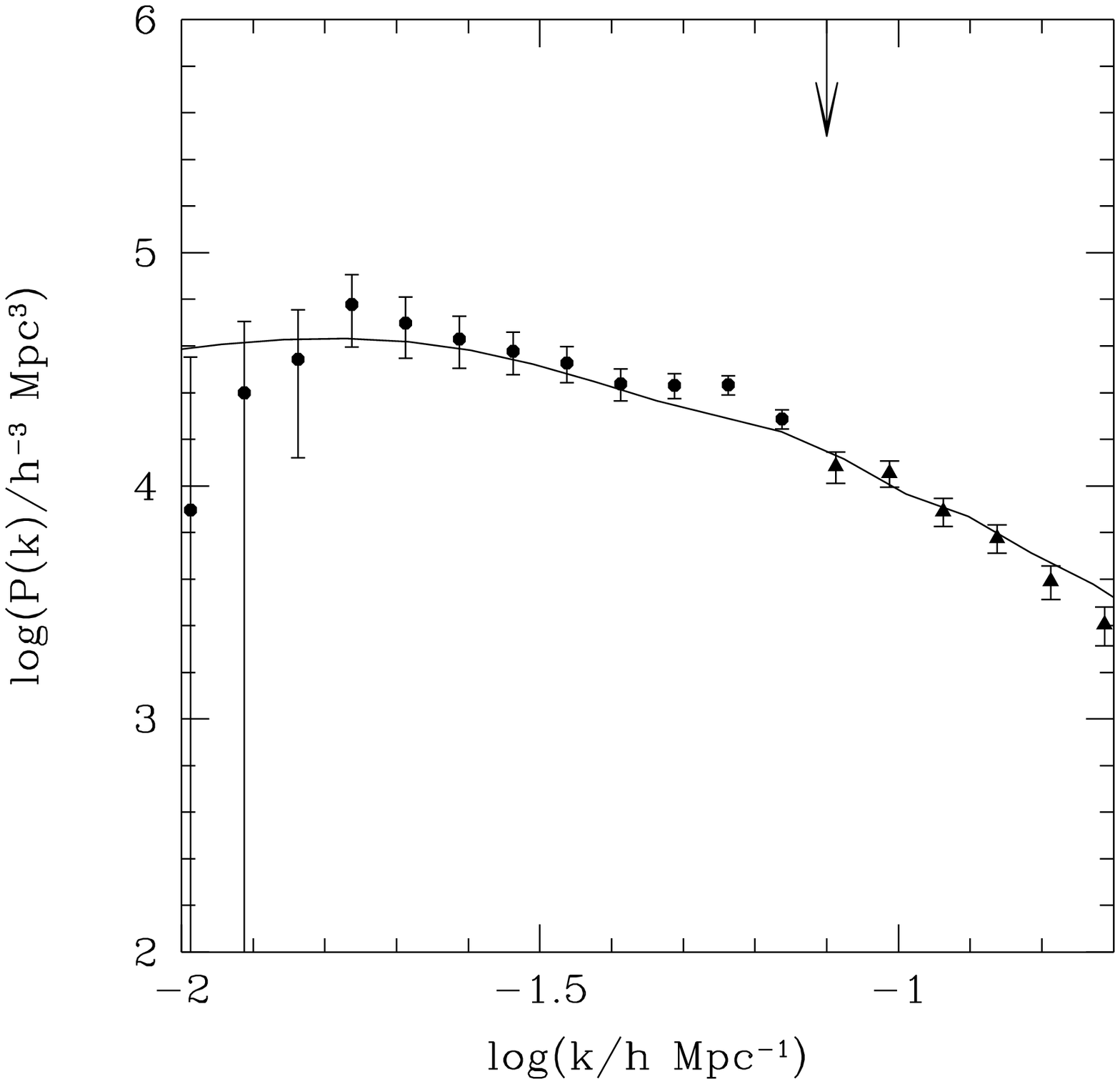}} \\
\end{tabular}
\caption{The circles in the left panel show the power spectrum from the
mock catalogues of the finished 2QZ found using a FFT. The
triangles in the middle plot show the 
power spectrum from the mock catalogues using the direct Fourier
Transform method. In the right plot, the circles show the mock
catalogue power spectrum, estimated using a FFT on scales above the
arrow and the direct method on scales below the arrow. In all cases,
the line shows the real space input power spectrum to the 
{\it Hubble Volume} simulation, renormalised to mimic the effects of
bulk motions on the power spectrum amplitude. 
The open symbols demonstrate the points affected by problems 
with each different method of power spectrum estimation. On small scales,
the open circles lie below the input power spectrum as these scales
are below the FFT limit. On large scales, the open triangles lie
above the input power spectrum as we have not accounted for the window
function normalisation. A combination of the two methods gives a good
match to the input power spectrum over a wide range of scales, as seen
in the right hand plot.}
\label{fig:pkmeth}
\end{centering}
\end{figure*}

The biasing prescription is described in full detail in
Hoyle (2000). We provide a brief summary here. We split the
simulation into slices of 0.2 in redshift. Within each slice, we bin
the redshift space density field onto a grid of cell size
20$h^{-1}$Mpc and calculate the mean density and standard deviation
for each cell. Following Cole et al. (1998), the bias probability is given by
\begin{equation}
P(\nu) =  \left\{ \begin{array}{ll}
   \exp(\alpha\nu + \beta\nu^{3/2}) & \mbox{if $\nu\geq0$} \\ 
    \exp(\alpha\nu) & \mbox{otherwise},
     \end{array} \right .
\label{eq:biasprob}
\end{equation}
where $\nu$ is the number of standard deviations of the cell density
away from the mean cell density. We fix $\alpha$ to be 0.15, similar
to the value given in Cole et al. such that the real space clustering
at present day in a $\Lambda$CDM simulation with parameters similar to
those of the {\it Hubble Volume}
approximately matches that of the APM galaxy survey (Baugh \& Efstathiou 1993).
We vary
$\beta$ until the correlation length of biased particles in each
redshift slice is consistent with 6$h^{-1}$Mpc over the range 
$5<r< 30 h^{-1}$Mpc, assuming that the correlation function has the
form $\xi(r) = (r/r_{\circ})^{-1.7}$.

As the clustering amplitude of the dark matter decreases with redshift
proportional to the square of the linear growth factor, the bias
factor required to keep the clustering amplitude constant with
redshift must correspondingly increase with redshift, proportional
 to the growth factor.  At the average redshift of the survey, the
bias factor, $b$, is approximately $\sqrt{4.7}$.
We then imprint the radial selection function of the QSOs 
onto the mock catalogues as described in Section \ref{sec:rancat}. 
 
The net result is shown in Figure \ref{fig:pkbias}. The points show
P(k) for the
biased particles, the solid line shows P(k) of the mass and the dashed
line shows $b^2$P(k)$_{\rm mass}$. In both cases, the
QSO selection function has been applied to the catalogues and the
redshift space distortions included in the particle
positions. By design, over a wide range of scales, there
is just a linear bias between the mass and the biased particles. 

In Figure \ref{fig:pkmeth}, we test the method of estimating the power
spectrum on the mock catalogues of the finished 2QZ. The solid
line shows the input power spectrum to the {\it Hubble Volume}
simulation. The power spectra from 2QZ and the mock catalogues are 
measured in redshift space,
whereas the input power spectra is calculated in real space. Small
scale peculiar velocities affect the shape of the redshift power
spectrum on small scales. However, they should have little effect
on the shape of the power spectrum on scales $\gsim 5 h^{-1}$Mpc (see
Hoyle 2000), which is below the scale where we can measure QSO
power spectra.

On large scales, the effect of redshift space distortions is to boost the 
amplitude of the power spectrum according to 
\begin{equation}
P_s(k) = P_r(k)\left(1 + \frac{2}{3}\beta + \frac{1}{5}\beta^2 \right),
\end{equation}
(Kaiser 1987) with $\beta = \Omega_{\rm m}^{0.6}/b$ and $r$ and $s$ 
indicating the real space and redshift space power spectra. The redshift 
space distortions are caused by bulk motions of QSOs. If the bias is scale 
independent the redshift space power spectrum just has a higher amplitude 
than the real space power spectrum but the shapes should be consistent.
Therefore we test our method of recovering the power spectrum by
comparing to the shape of the input power spectrum.

The circles in the left plot show the power spectrum from the mock 
catalogues, averaged over the
three realisations, estimated using a FFT. On intermediate to large scales, 
(60-400$h^{-1}$ Mpc or $-1.8< {\rm log}(k) <-1$) the power
spectrum from the mock catalogues using the FFT reproduces the shape
of the input power spectrum very well. On the very largest scales, 
the points are systematically below
the line due to the effects of the window function. A convolution of the input 
power spectrum with the window function power spectrum reveals a very similar
decrease in power on these large scales (Hoyle 2000). 
It is very difficult to deconvolve fully the window function from the
true power spectrum. Instead of attempting this, we estimate the scale on
which the window function affects the shape of the power spectrum
through comparisons with the input power spectrum. Assuming the
$\Lambda$ cosmology, the power spectrum will be free from any effects
of the window function on scale less than 400$h^{-1}$Mpc 
(log($k / h$Mpc$^{-1}$)=$-1.8$). 
On small scales,
the circles lie below the input power spectrum. These scales are below
the limit of the FFT and the power spectrum cannot be robustly
measured.

Instead, on small scales we use a direct Fourier transform to
calculate the power spectrum. This is shown by the triangles in the
middle plot. The solid
triangles reproduce the input power spectrum well on scales smaller than 
$\approx$100$h^{-1}$Mpc (log($k / h$Mpc$^{-1}$)=$-1.2$) where the window 
function has
little effect on the amplitude of the power spectrum. On larger
scales, the effect of not subtracting off the window function in the
estimation of the power spectrum can
clearly be seen. 

In the right hand plot, we show a combination of the two methods. On
scales larger than the arrow, we use the FFT method and on scales
smaller than the arrow, we use the direct method. This enables a robust 
measurement of the power spectrum over a wide range of scales.
The scale at which we swap the method of estimating the power spectrum is
the central point at which the two methods of estimating the power spectrum 
agree within the 1$\sigma$ errors.

\begin{figure*} 
\begin{centering}
\begin{tabular}{cc}
{\epsfxsize=8truecm \epsfysize=8truecm \epsfbox[35 170 550 675]{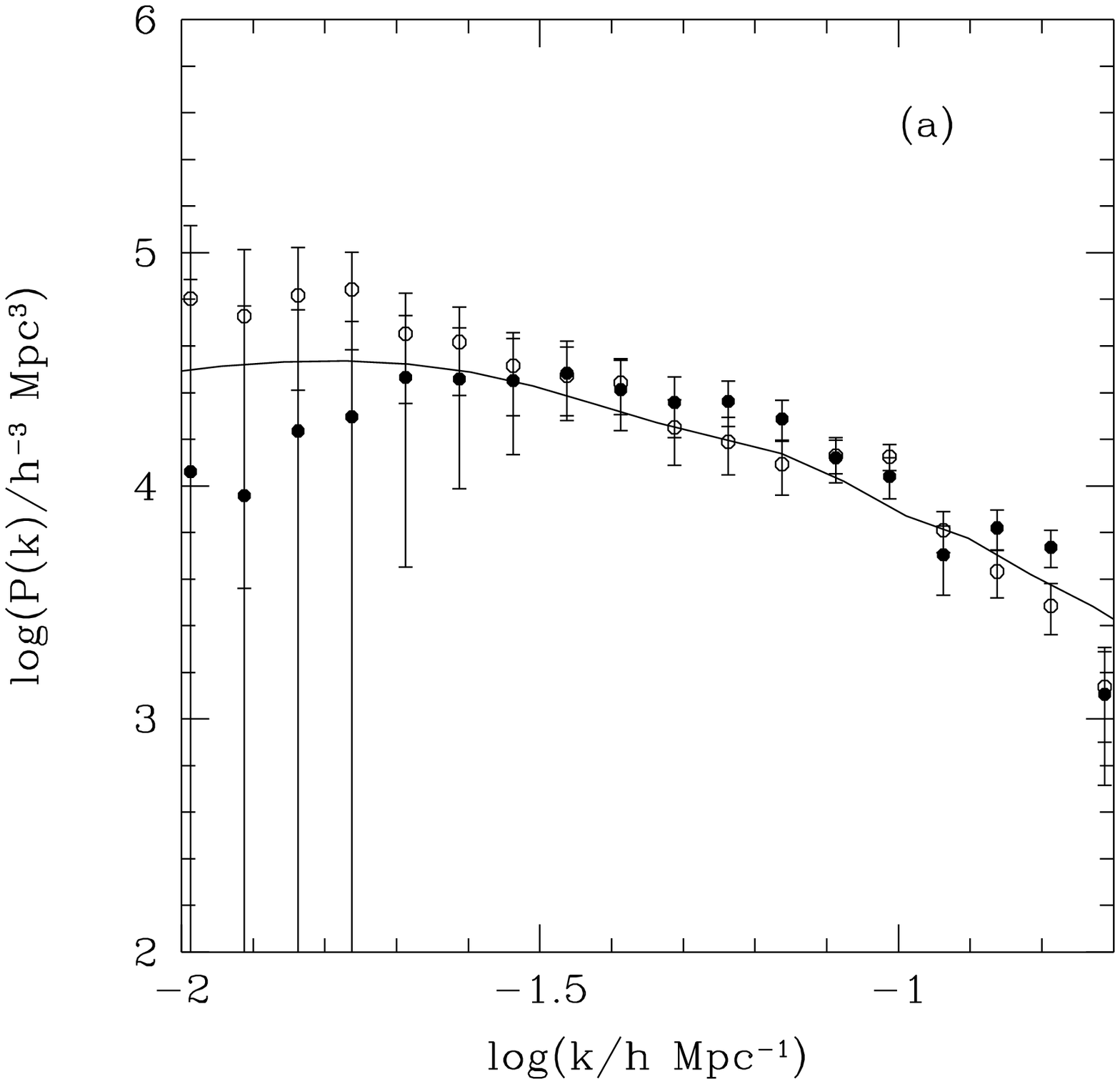}} & 
{\epsfxsize=8truecm \epsfysize=8truecm \epsfbox[35 170 550 675]{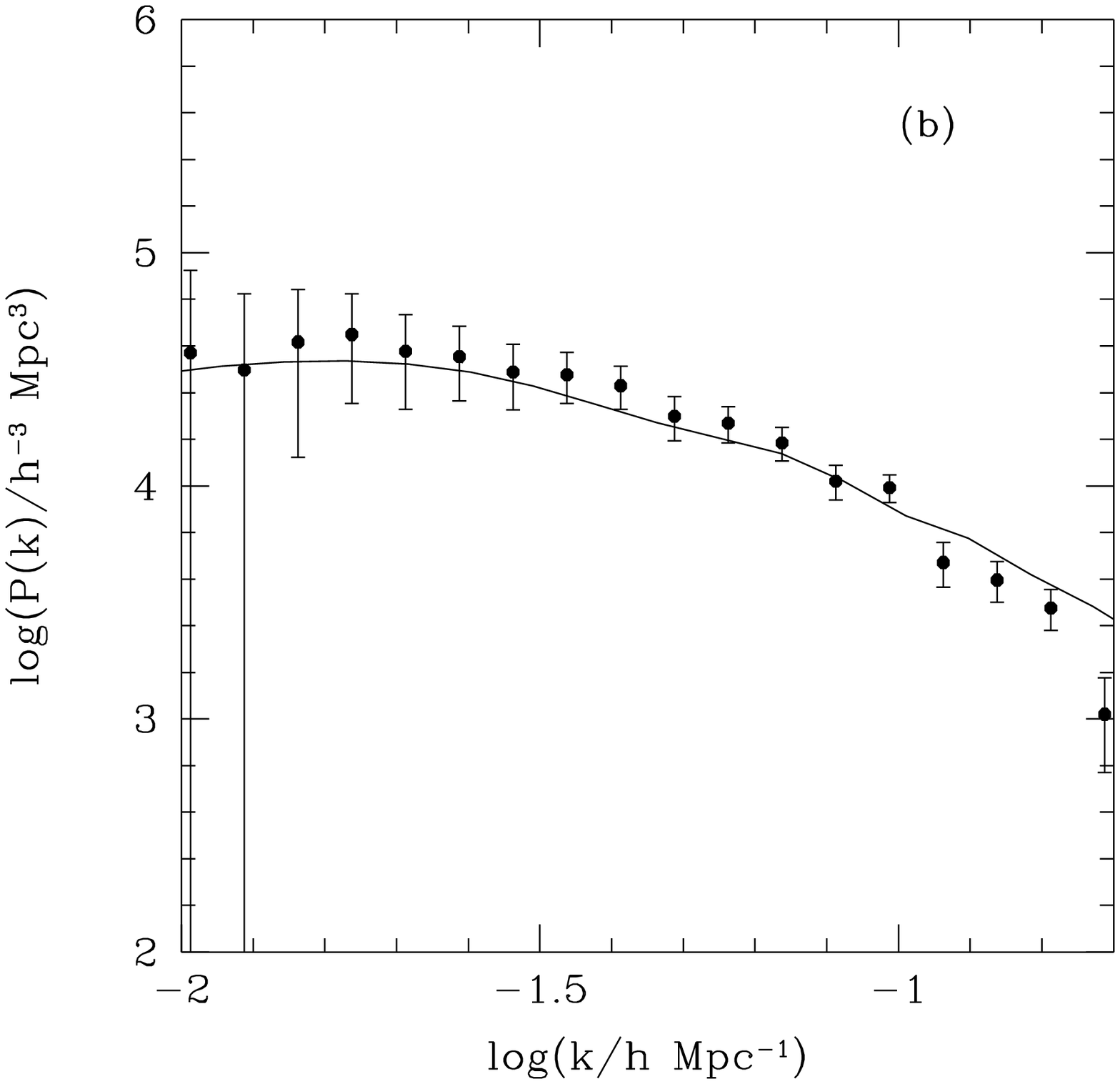}} \\
\end{tabular}
\caption
{The points in the left hand plot show the power spectra measure from
mock SGC simulations (open circles) and mock NGC simulations (closed
circles). These are combined in the right hand plot, as described in
the text, to give a single mock power spectrum with the current
angular selection function imprinted on the mock catalogues. The solid
line in both plots is the input power spectrum, as 
described in Figure \ref{fig:pkmeth}. The
power spectrum measured with the current angular selection function
imprinted matches the input power spectrum over a wide range of scales.
In both plots, the errors are FKP errors and we use a FFT to measure the
power spectra on scales larger than 70$h^{-1}$Mpc
(log($k/h$Mpc$^{-1}$)=$-1.12$) and the direct method on smaller
scales. The simulation has the $\Lambda$ cosmology.}
\label{fig:pkgp}
\end{centering}
\end{figure*}

The 2QZ is, however, not yet finished and currently has a patchy
angular selection function, as discussed in section \ref{sec:rancat_ang}. To
test the effect of this on the power spectrum, we imprint the
angular selection function of the NGC and SGC onto the two outer mock 
catalogues, as described in Section \ref{sec:rancat_ang}. 
Figure \ref{fig:pkgp}(a) shows the average power spectrum after the
angular selection function of the NGC (filled
circles) and the SGC (open circles) has been imprinted onto the mock 
catalogues. 

In Figure \ref{fig:pkgp}(b), we combine the power spectrum of the NGC
and SGC together, weighted by the inverse of the variance, to create
one single mock 2QZ power spectrum. The same combination of FFT and 
direct Fourier Transform, as described above, 
was used in the calculation of the power spectrum. 

Over the range of values $-1.8 < {\rm log}(k) < -1$ (60-400$h^{-1}$Mpc), 
the power spectrum from the NGC and
SGC mock catalogues and the combined power spectrum 
match the input power spectrum reasonably well
although there is, perhaps, a slight steepening of the power spectrum. 
On larger and smaller scales, the agreement between the mock power
spectra and the input power spectrum worsens compared to the agreement
between the mock, finished power spectrum and the input power
spectrum. This is due to the smaller
volume of the incomplete survey, the smaller number of QSOs
currently observed and errors that, through tests on the 
{\it Hubble Volume} (see Section \ref{sec:errors} and Hoyle 2000), 
are perhaps underestimated using the FKP method.

The tests show that even with the incomplete 2QZ, accurate
measurements of the power spectrum can be made, although over a
slightly limited range of scales as compared to the power spectrum
predicted from the final survey, with errors that are larger (by a
factor of $\approx 1.5$) than the finished survey will have.

\section{Results}
\label{sec:qsoreslts}

\begin{figure*} 
\begin{centering}
\begin{tabular}{cc}
{\epsfxsize=8truecm \epsfysize=8truecm \epsfbox[35 170 550 675]{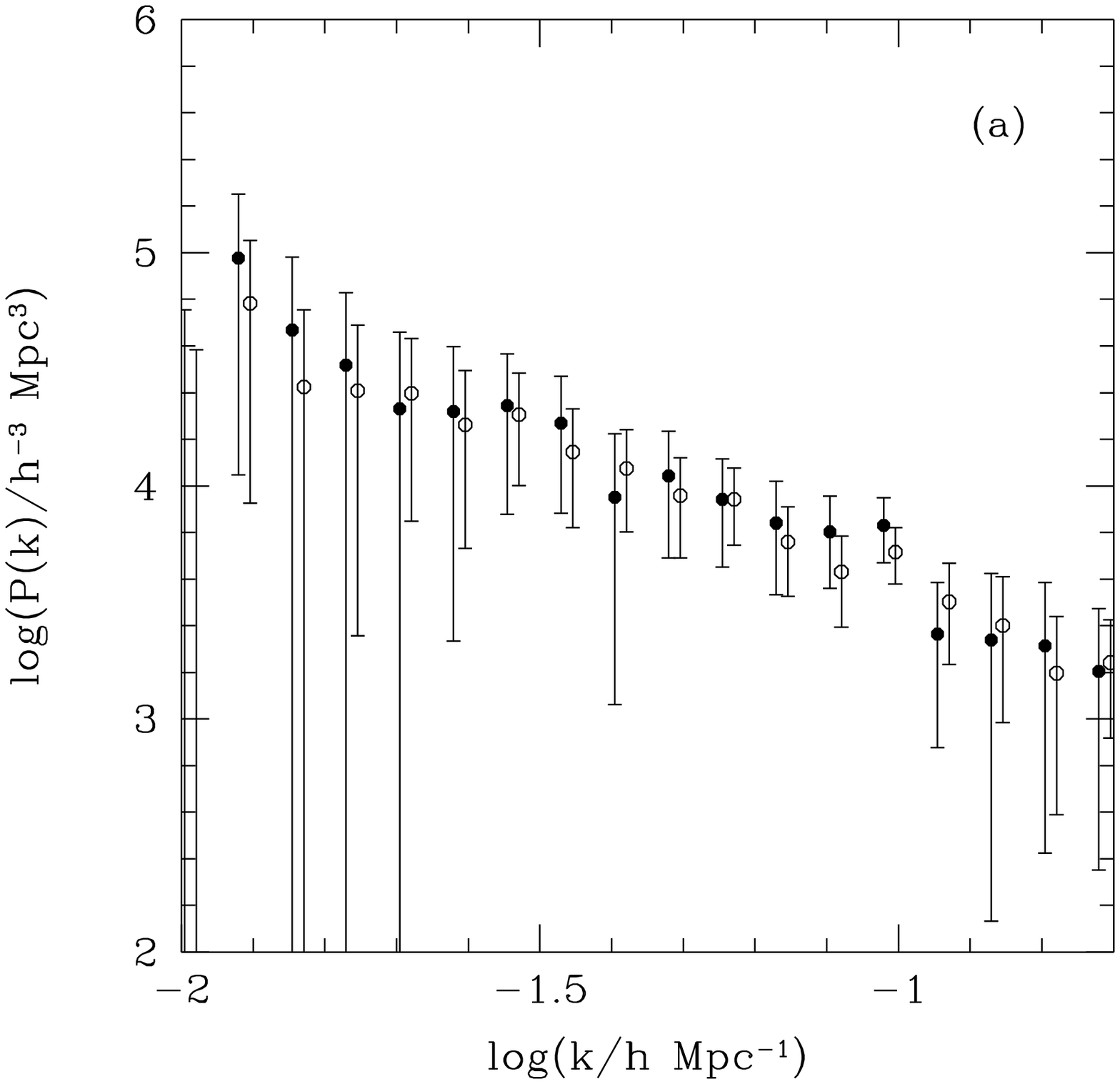}} &
{\epsfxsize=8truecm \epsfysize=8truecm \epsfbox[35 170 550 675]{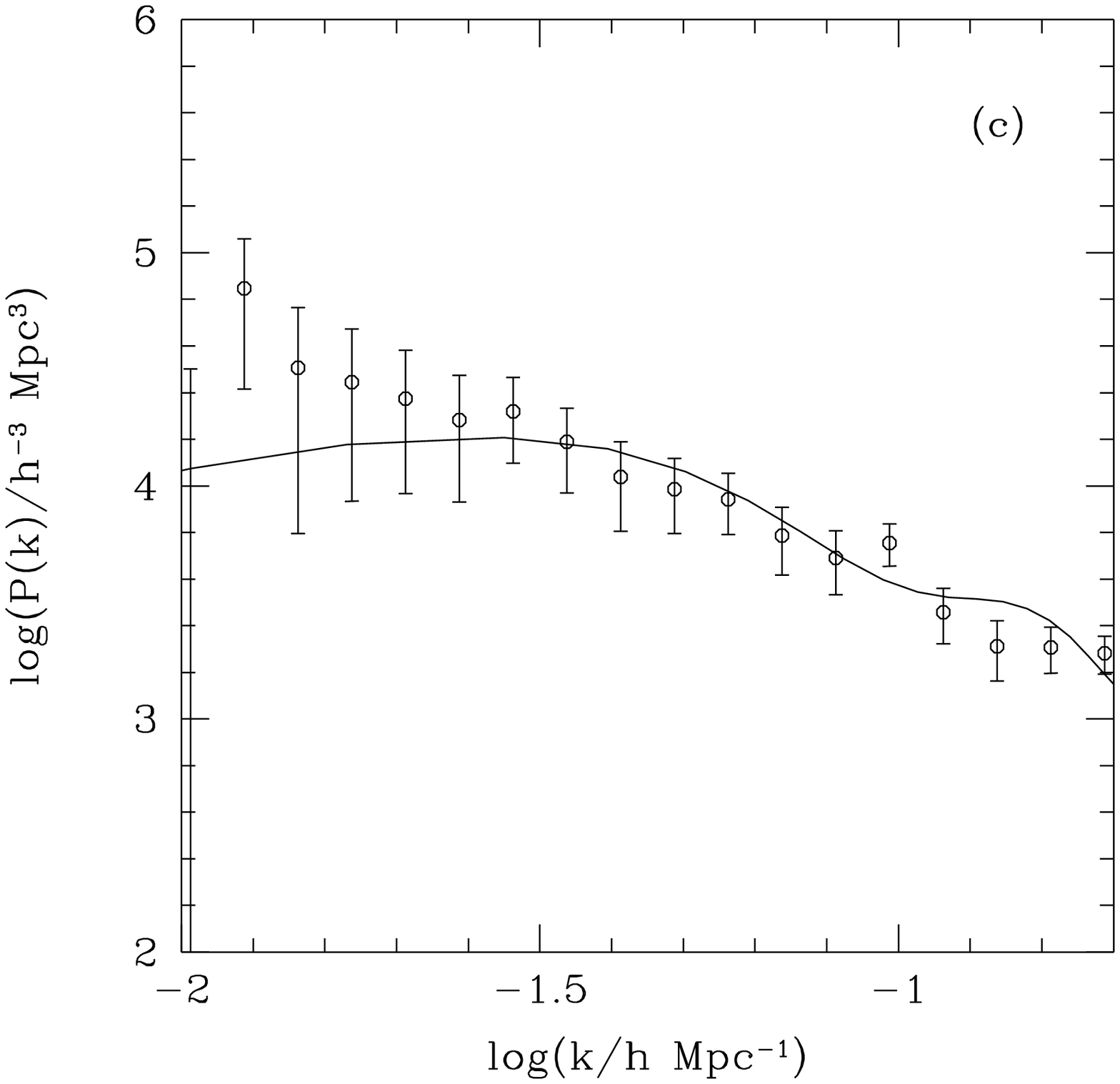}} \\
{\epsfxsize=8truecm \epsfysize=8truecm \epsfbox[35 170 550 675]{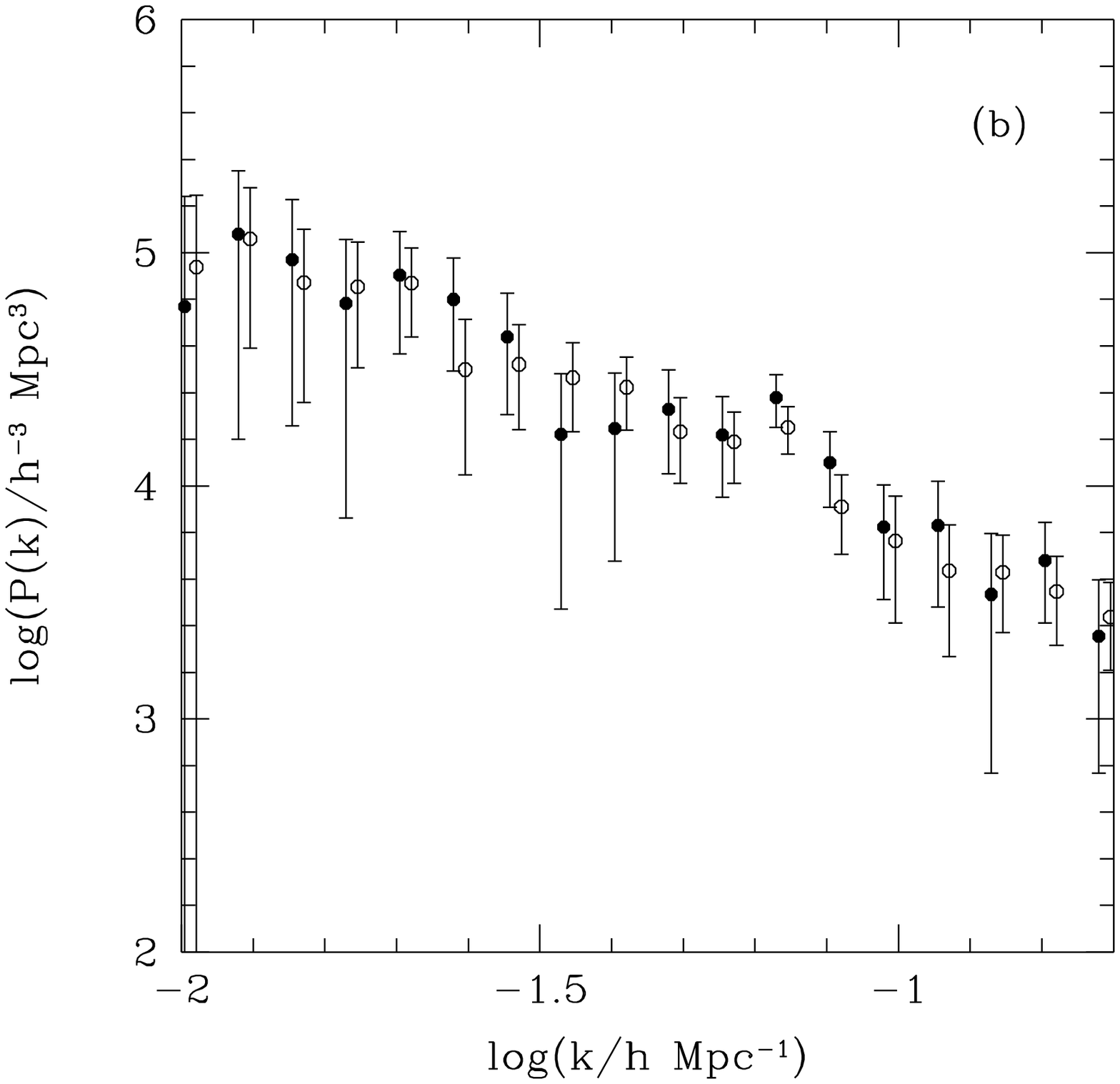}} &
{\epsfxsize=8truecm \epsfysize=8truecm \epsfbox[35 170 550 675]{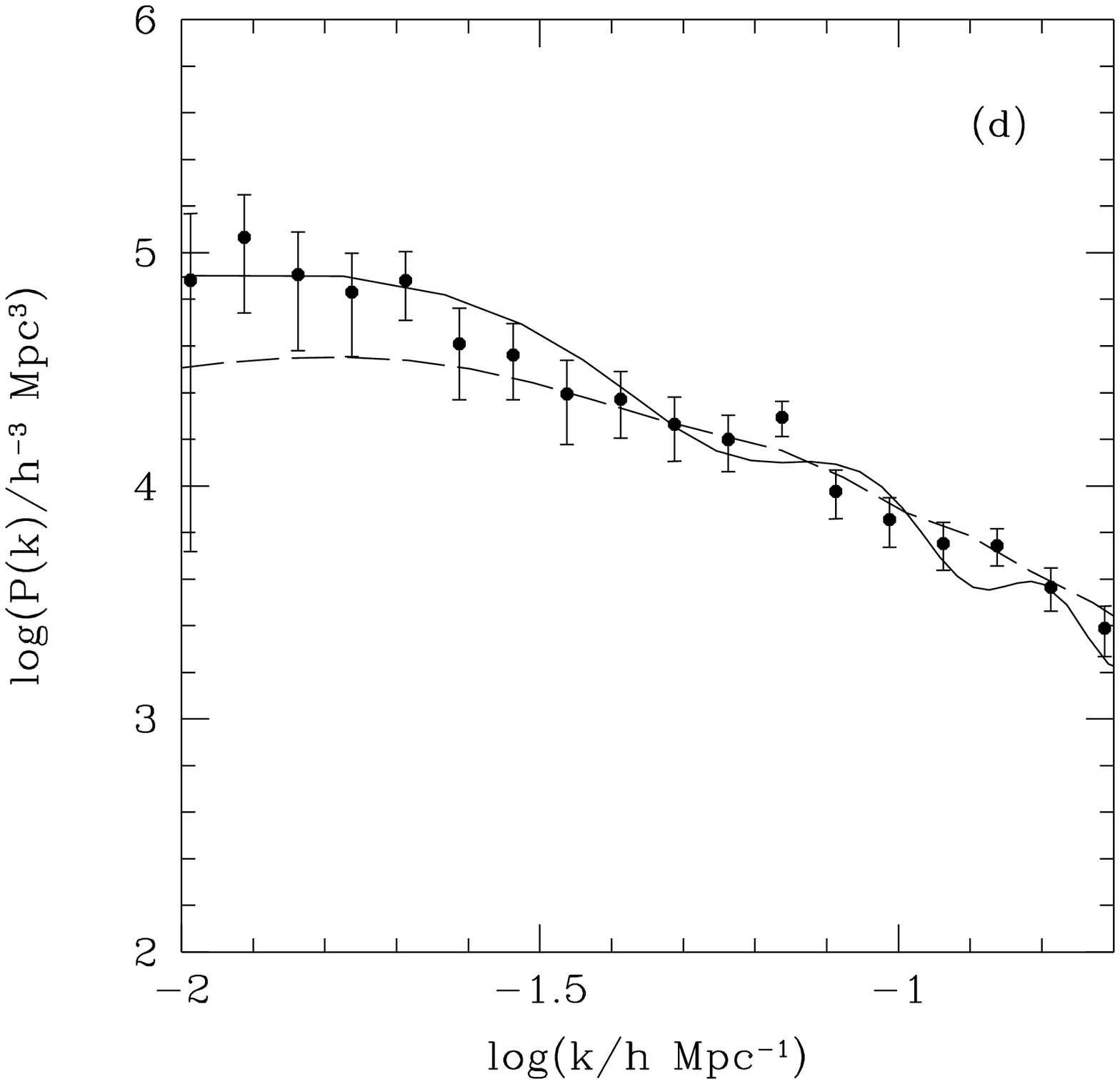}} \\
\end{tabular}
\caption {In Figures (a) and (b) we compare the
power spectrum from the NGC (filled circles) and SGC (open circles). 
Both sets of 
points have been offset by 0.01 in log$(k/ h {\rm Mpc}^{-1})$ for clarity. In
Figures (c) and (d) we show the combined QSO power spectra. 
The EdS cosmology is assumed for (a) and (c) and the $\Lambda$
cosmology for (b) and (d). The solid lines in (c, d) show models with 
an increased baryon fraction (described in the text) to try and reproduce 
the feature present in the QSO power spectra. The dashed line in (d) is the 
input power spectrum to the {\it Hubble Volume} simulation.
QSOs have redshifts in the range $0.3<z<2.2$. }

\label{fig:pk_ns}
\end{centering}
\end{figure*}

\subsection{The QSO Power Spectra}
\label{sec:gamma}

In this Section, we present QSO power spectra measured from the NGC,
SGC as well as the combined power spectra for the two choices of
cosmology. We measure the power spectrum from QSOs with redshifts in 
the range $0.3<z<2.2$ where the photometric completeness is higher than
$\approx$90 per cent Boyle (2000) and the number density remains 
approximately constant.

In Figure \ref{fig:pk_ns} we compare the NGC and SGC power spectra 
measured from QSOs with redshift in the range $0.3<z<2.2$ assuming
the EdS cosmology (a) and the $\Lambda$ cosmology (b).
The combined power spectra are shown in (c, EdS cosmology) and (d,
$\Lambda$ cosmology). There is good agreement between the power 
spectrum measured from the NGC and the SGC strips over a wide range of scales.
The errors we show here and through out are 1$\sigma$ errors.

The power spectra have a power-law form with $P_{\rm QSO}(k)\propto k^{\alpha}$
with $\alpha \sim -1.3$ for both choices of cosmology 
(-1.26$\pm0.15$ EdS and -1.31$\pm0.16$ $\Lambda$) over the range
$-1.7 < {\rm log}(k) < -1$ (60-300$h^{-1}$Mpc).

CDM models predict that the power spectrum should turn over at a scale 
which depends on the shape parameter, $\Gamma$, of the power spectrum. In 
pure CDM models with a scale-invariant spectrum of initial fluctuations 
$\Gamma=\Omega_{\rm m}*h$. This relation becomes more complicated, however, 
with the 
addition of baryons, hot dark matter, or a tilted initial fluctuations 
spectrum, and so a measurement of Gamma cannot be reliably inverted to 
imply a measurement of $\Omega_{\rm m}*h$ (Sugiyama 1995, Peacock 2000).
If $\Gamma > 0.25$ then the turnover should occur on scales less than 
300$h^{-1}$Mpc. The fact that we do not see a turnover puts 
constraints on the
range of acceptable $\Gamma$ values. In Section \ref{sec:qsoimp}, we
compare the data to models to constrain the value of $\Gamma$ further.

\begin{figure*} 
\begin{centering}
\begin{tabular}{c}
{\epsfxsize=16truecm \epsfysize=16truecm \epsfbox[35 170 550 675]{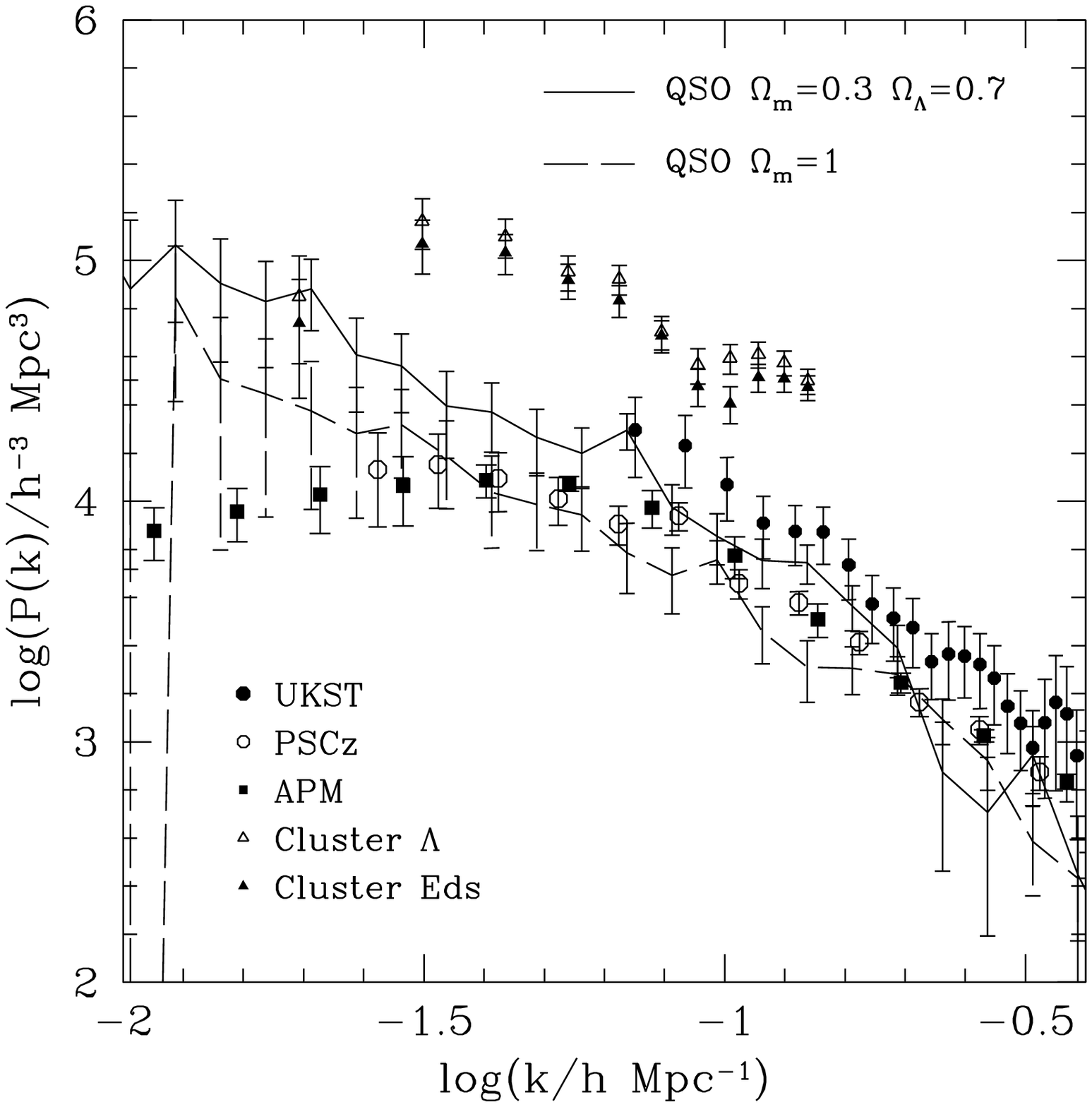}} \\
\end{tabular}
\caption {A comparison of galaxy and QSO power spectra. The lines show the QSO power spectra with the $\Lambda$ cosmology (solid line) and the EdS cosmology (dashed line) assumed. The filled circles show a flux limited power spectrum measured from the Durham/UKST Survey (Hoyle et al. 1999), the open circles show the power spectrum measured from the PSCz Survey (Sutherland et al. 1999). The filled squares show the APM real space galaxy power spectrum taken from Baugh \& Efstathiou (1994). The triangles show the power spectrum of rich clusters with $\Omega_{\rm m}$=1 (filled) and $\Omega_{\rm m}$=0.2, $\Omega_{\Lambda}$=0.8 (open) taken from Tadros, Efstathiou \& Dalton (1998). The errors on the QSO power spectra are 1$\sigma$ FKP errors, the errors on the Durham/UKST and PSCz galaxy power spectrum and the errors on the cluster power spectrum are from mock catalogues. The errors on the APM power spectrum are obtained by measuring the scatter in power spectra obtained from four separate regions of the APM survey.}
\label{fig:pscomp}
\end{centering}
\end{figure*}

We tentatively identify a `spike' feature at 
log($k / h$Mpc$^{-1}$) $\approx -1.16$ (90 $h^{-1}$Mpc) assuming the $\Lambda$ 
cosmology and at log($k / h$Mpc$^{-1}$) $\approx -1.01$ (65 $h^{-1}$Mpc) 
assuming the EdS. Figures \ref{fig:pk_ns}(a,b) show that the 
feature seems to reproduce in the NGC and SGC strips. We shall see in
Figure \ref{fig:pkevo}(a,b) that this feature also appears to reproduce in the
two independent redshift bins in both cosmologies.
In both cases, the spike feature lies about 2$\sigma$ above the powerlaw 
fit to the power spectrum. (slightly higher in the $\Lambda$ case, slightly 
lower in the EdS case), corresponding to a probability of 1/20 that a point 
deviates from the fit, neglecting the fact that the points may not be 
completely independent. With 17 data points, we should expect one point
to deviate by this amount so the spike feature is marginal.
However, if real, such a feature might be caused by
acoustic oscillations in the baryon-radiation fluid prior to decoupling. If
this power spectrum spike is confirmed in the analysis of the full 2QZ, it would
also
provide a powerful measure of the world model. Such features do not evolve in
scale as a function of redshift and a comparison between the 2QZ P(k), measured
at a mean redshift of 1.4 and the galaxy P(k) measured at z$\approx$0 may allow
constraints on $\Omega_{\rm m}$ and $\Omega_\Lambda$ via the 
requirement to maintain
the feature at the same scale in both power spectra. 
For example, preliminary reports of results from the 2dF Galaxy Redshift
Survey power spectrum (Peacock et al. 2001) show some possible indications
of a spike detected in their data at log($k/ h {\rm Mpc}^{-1}$) $\approx$
-1.12 (85$h^{-1}$Mpc), assuming the $\Lambda$ cosmology, although the 
2dFGRS team do not claim that any single isolated feature in their data is
significant. In an EdS cosmology this feature would lie at log($k/ h {\rm 
Mpc}^{-1}$) $\approx$ -1.08 (75$h^{-1}$Mpc).
The low redshift galaxy result may therefore be somewhat more
consistent with the high redshift QSO result assuming the
$\Lambda$ cosmology but more QSO and galaxy data will clearly be needed
at both high and low redshift before a final conclusion can be drawn.

We have used the CMBFAST programme (Seljak \& Zaldarriaga 1996)
to see how close the feature might be in scale and amplitude to the
second baryon-radiation acoustic peak. Generally, for a wide range of
$\Omega_b$ and $h$ in both the EdS and the $\Lambda$ cosmology, 
the peak scale appears at $\gsim$25 per cent smaller wavenumber than the 
predicted second peak. Also
the amplitude of the peak appears to demand a higher value of $\Omega_b$
than used in e.g. the $\Lambda$CDM Hubble Volume model ($\Omega_\Lambda$=0.7, 
$\Omega_{\rm m}$=0.3 $\Omega_b$=0.04, $h$=0.7). To illustrate this, the 
{\it Hubble Volume} input spectrum is shown in Figure \ref{fig:pk_ns}(d) 
together with a CMBFAST
prediction with $\Omega_\Lambda$=0.7, $\Omega_m$=0.3, $\Omega_b$=0.1, $h$=0.5. 
This model gives as good a fit as any to the observed spike feature but, 
although having a reasonable amplitude, it lies at 
log$(k/h {\rm Mpc}^{-1}$)=-1.05 rather than
the observed log$(k/h {\rm Mpc}^{-1}$)=-1.16. 
Figure \ref{fig:pk_ns}(c) shows a CMBFAST model 
with $\Omega_m$=1,
$\Omega_b=0.3, h$=0.3 again as an example of a model which gets as close as any 
to matching the observed feature. However, again the predicted spike lies
at too large a wavenumber, log($k/h {\rm Mpc}^{-1}$)=-0.82, compared to 
the observed spike 
scale of  log($k/h {\rm Mpc}^{-1}$)=-1.02. 
We also note that Meiksin, White \& Peacock (1999) have
suggested that quasi-linear evolution  may reduce the predicted amplitude 
of the second acoustic peak.
We postpone further discussion of the theoretical implications of the spike 
feature until we see if it reproduces in our full 25k 2QZ sample.

\subsection{Comparison with P(k) from Galaxy and Cluster Surveys}

The power spectrum of optically selected galaxies has been 
measured from several different galaxy surveys but so far only out to 
$\approx 100 h^{-1}$Mpc scales. In Figure \ref{fig:pscomp}, we compare the power 
spectrum estimated from the QSOs currently observed in 2QZ, assuming two different 
cosmologies, with a power spectrum from the Durham/UKST survey 
(Hoyle et al. 1999). The 
lines show the QSO power spectrum, estimated assuming the $\Lambda$ cosmology 
(solid line) and the EdS cosmology (dashed line). The filled circles show a flux 
limited galaxy power spectrum with P=8000$h^{-3}$Mpc$^3$ fixed in the weighting 
scheme of Feldman, Kaiser \& Peacock (1994). 

Figure \ref{fig:pscomp}  shows that the form of the QSO and galaxy redshift space 
power spectra are similar, although the range of overlap where the two power 
spectra are robustly measured is small. The slopes are 
statistically consistent whether the QSO power spectrum is computed in the 
EdS or $\Lambda$ cosmology. In the final
sample there may be a possible cosmological test afforded by
comparing the slopes of QSO power spectra with power spectra
obtained at lower redshift, if the bias can be assumed to be scale
independent. The QSO power spectrum in the $\Lambda$ case has a
amplitude slightly lower than that of the galaxy power spectra. 
If an EdS cosmology is
assumed, the amplitude of the QSO power spectrum (dashed line) is
approximately a factor of $\approx$3 lower than the galaxy power
spectra. Croom et al. (2001) found very
similar results in the comparison of galaxy and QSO clustering   
amplitudes in their correlation function analysis of the 2QZ data.

We compare the QSO power spectra to the PSCz Survey power spectrum 
(Sutherland et al. 1999) in 
Figure \ref{fig:pscomp}. The PSCz Survey is an Infra-Red selected redshift survey. 
Comparisons between optically and IR selected surveys reveal that there is 
typically an offset between clustering statistics between the two surveys with 
P$_{\rm Opt} = 1.3^2$ P$_{\rm IR}$ (Hoyle et al. 1999) which explains why 
the PSCz P(k) 
has an amplitude that is in closer agreement with the EdS QSO P(k) than the 
$\Lambda$ P(k), contrary to the optical comparison.

We also compare the QSO power spectra to the real space APM galaxy
power spectrum of Baugh \& Efstathiou (1994), inferred from measurements of the
angular correlation function, $w(\theta)$. 
The real space APM galaxy power spectrum has a lower
amplitude than the Durham/UKST redshift space power spectrum as 
bulk motions of galaxies increase amplitude of the redshift space 
power spectrum (Kaiser 1987). There is also some
uncertainty in the normalisation of the APM power spectrum due to the 
inversion technique of estimating the power spectrum 
(Carlton Baugh, private communication).
On scales out to 100 $h^{-1}$Mpc (log($k / h$Mpc$^{-1}$)=$-1.2$), there is
reasonable agreement between the amplitudes of the different power
spectra. 
On scales larger than 100$h^{-1}$Mpc, the APM power spectrum flattens
off, whereas the QSO power spectra continue to rise. We believe that the
QSO power spectra measurements are robust to 300$h^{-1}$Mpc assuming the 
EdS cosmology (400$h^{-1}$Mpc assuming the $\Lambda$ cosmology). Therefore
we find even more large scale
power than in the APM power spectrum and possibly even more power than
the $\Lambda$CDM models (see Section \ref{sec:qsoimp}).

A final comparison is made between the QSO power spectra and the power
spectra of rich clusters of galaxies. In Figure \ref{fig:pscomp},
the triangles show
the APM Cluster Survey power spectra taken from
Tadros, Efstathiou \& Dalton (1998). 
The cluster power spectra are found assuming
$\Omega_{\rm m}$=1 (filled triangles) and 
$\Omega_{\rm m}$=0.2, $\Omega_{\Lambda}$=0.8 (open triangles). As the
clusters are only observed out to a redshift of $\approx$0.2, the
effects of cosmology on the clustering amplitude are small so the two
cluster power spectra have a similar amplitude and shape. The slopes
of the QSO power spectra are similar to those for the clusters for
both models, within the overlapping range of scales. As can clearly be
seen in Figure \ref{fig:pscomp}, the QSOs have a lower clustering
amplitude than the present day rich clusters. We shall see below that
the QSO power spectrum amplitude only evolves slowly with redshift and 
so it is meaningful to compare the QSO and cluster power spectra.
The relative bias
between the two types of power spectra is $b_{\rm rel} \approx \sqrt{5}$
for the $\Lambda$ cosmology and $b_{\rm rel} \approx \sqrt{10}$ for the
EdS cosmology.

\subsection{The Evolution of QSO Clustering}
\label{sec:pkevo}

\begin{figure} 
\begin{centering}
\begin{tabular}{c}
{\epsfxsize=8truecm \epsfysize=8truecm \epsfbox[35 170 550 675]{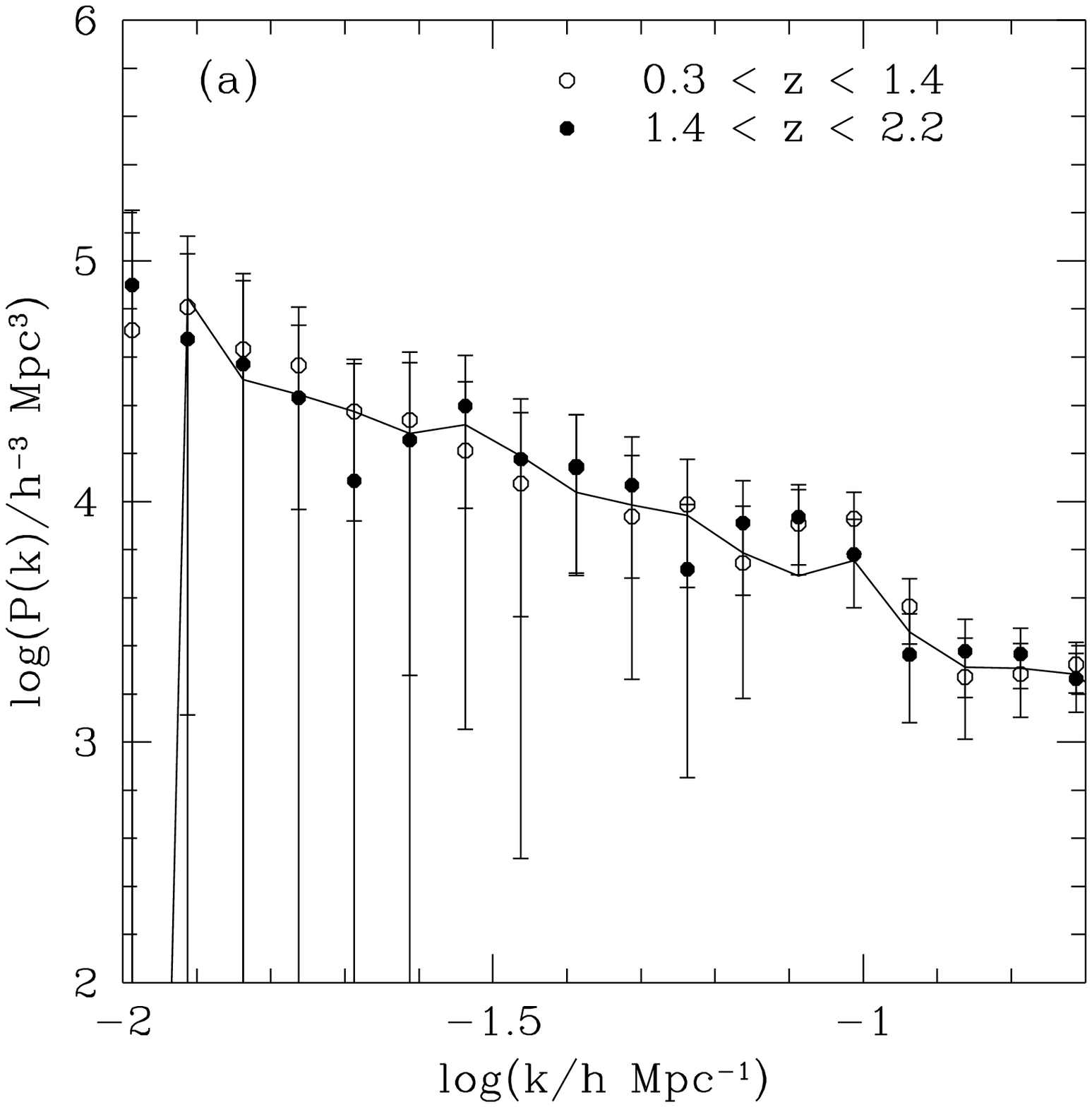}} \\
{\epsfxsize=8truecm \epsfysize=8truecm \epsfbox[35 170 550 675]{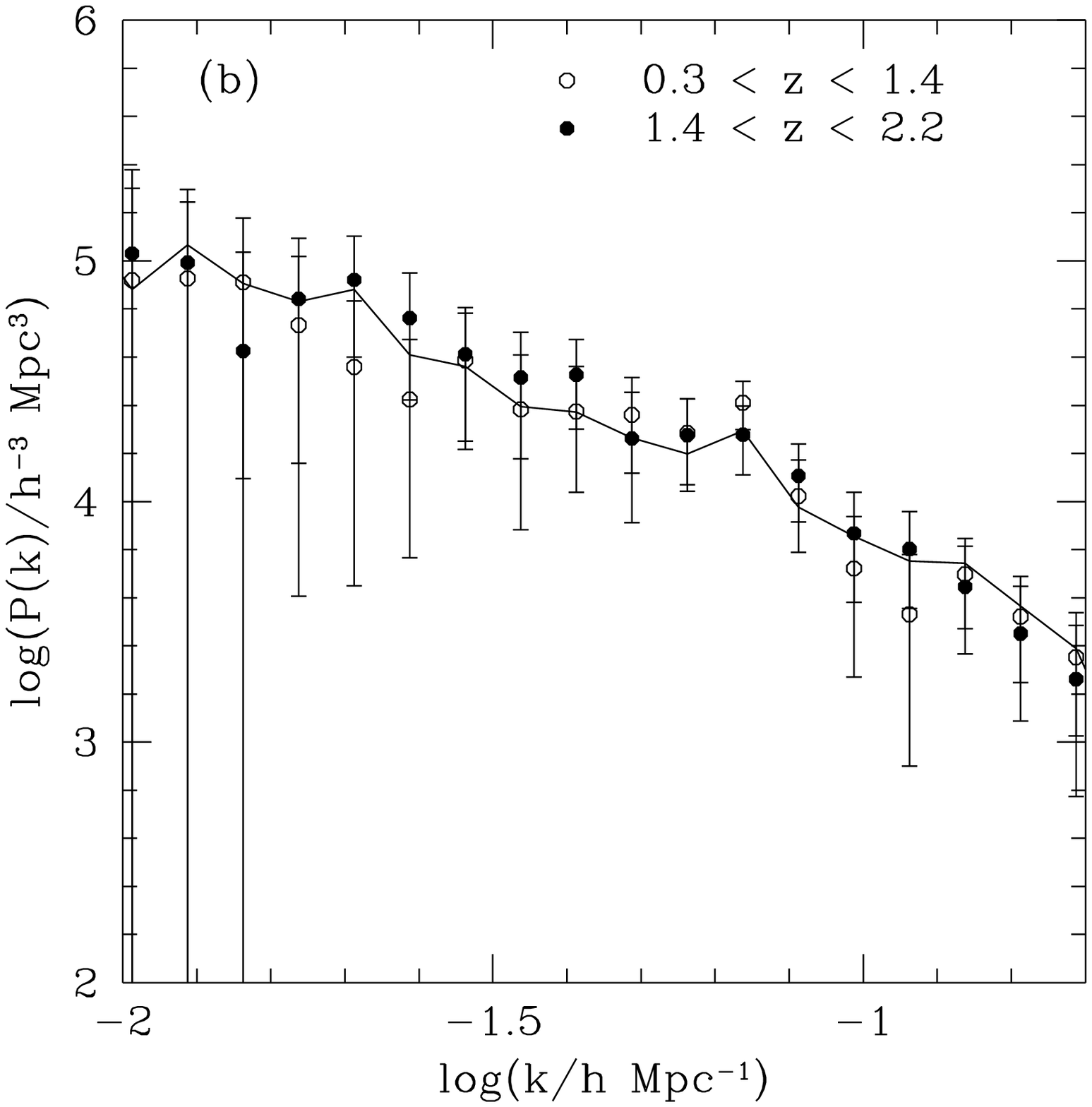}} \\
\end{tabular}
\caption{The power spectrum of QSOs measured at different redshifts. The EdS cosmology is assumed in (a) and the $\Lambda$ cosmology is assumed in (b). In both panels, the solid line shows the power spectrum estimated from all the QSOs with $0.3 < z < 2.2$, as in Figure \ref{fig:pkpred}. The open circles show the power spectrum of QSOs with redshifts in the range $0.3<z<1.4$ and the filled circles show the power spectrum of QSOs with redshifts in the range $1.4<z<2.2$. The errors are 1$\sigma$ FKP errors. The points are slightly offset for clarity.}
\label{fig:pkevo}
\end{centering}
\end{figure}

To determine how QSO clustering evolves as a function of redshift, we
split the QSOs in each strip of the survey into two redshift bins
containing roughly equal numbers of QSOs, one with QSOs in the range
$0.3<z<1.4$ ($\bar{z}=1.0$) and the other with $ 1.4<z<2.4$
($\bar{z}=1.8$). The power spectrum is then measured from each
subsample on each strip, assuming the two cosmologies discussed in
Section \ref{sec:qsodata}. The results from the NGC and SGC for each
redshift bin and for each cosmology are averaged together as
before. All the power spectra in Figure \ref{fig:pkevo}(a) are
estimated assuming the EdS cosmology, whilst in panel (b), the
$\Lambda$ cosmology is assumed. In both panels, the solid line shows
the power spectrum of all the QSOs, the open circles show the power
spectrum of the low redshift QSOs and the filled circles show the
power spectrum of the high redshift QSOs.

When the QSOs are split up into the two samples, the errors on the
measured power spectra are increased. However, on scales around
100-200$h^{-1}$Mpc (-1.2 $>$ log($k / h$Mpc$^{-1}$) $>-1.5$)
the power spectrum can still be fairly well
measured, the fractional errors are around 40 per cent. There is
reasonable agreement between the power spectra measured from the two
redshift bins over a wide range of scales. In order to compare the amplitudes
of high and low redshift power spectra, we assume that they are both power
laws of the form log(P$/h^{-3}$Mpc$^{3}$) = $\alpha$log($k/ h$Mpc$^{-1}$) + A. 
We fix $\alpha$ to be -1.3 and find
the best fitting value of the amplitude, A, over the range 
-1.6$<$log($k/ h$Mpc$^{-1})<$-1.2 in the case of the 
$\Lambda$ cosmology and -1.5$<$log($k/ h$Mpc$^{-1})<$1.1 for the EdS case 
(corresponding to scales of 100-250$h^{-1}$Mpc for the $\Lambda$ case 
and 75-200$h^{-1}$Mpc in the EdS case). In the $\Lambda$ cosmology, the high 
redshift bin has an amplitude of 2.58$\pm{0.09}$ and the low redshift
bin has an amplitude of 2.63$\pm0.08$. In the EdS case, the high
redshift bin has an amplitude of 2.31$\pm0.09$ and the low
amplitude bin has an amplitude of 2.29$\pm0.10$. 
This shows that QSO clustering does not evolve strongly with 
redshift as the low and high redshift bins have power law fits that
are consistent in amplitude.

This result is consistent with the work of Croom et al. (2001)
who have
measured the clustering from the 10k QSO catalogue in five bins of redshift. If the EdS cosmology is assumed then little evolution is seen in the clustering amplitude of QSOs, consistent with Croom \& Shanks (1996). If the $\Lambda$ cosmology is assumed, slow evolution is seen in the clustering amplitude of QSOs, with QSOs at high redshift having a slightly higher clustering amplitude than QSOs at low redshift.

The slow evolution in QSO clustering was interpreted previously in
terms of cosmology and the
effects of QSO bias (see, for example, Croom \& Shanks 1996). 
The slow evolution either suggests that $\Omega_{\rm m}$ is low, less than
0.1, so that the dark matter clustering does not evolve strongly with
redshift, or if $\Omega_{\rm m}$ is fairly high,
$\gsim 0.1$, then bias evolves as a function of redshift to
counteract the decrease in amplitude of the dark matter
clustering. Croom et al. (2001) make a detailed comparison between
the evolution of
QSO clustering and various models for the QSO-mass bias.
Determining cosmology from the evolution of QSO clustering
alone is not possible due to the degeneracy between cosmology and
bias.

Since the QSO power spectra have comparable amplitudes to local galaxy
power spectra and since the slow QSO clustering evolution implies that
the QSO amplitude does not increase markedly at low redshift, this 
suggests that QSOs are not 
highly biased objects with respect to present day optically selected
galaxies.

\section{Comparison with Models of Large Scale Structure}
\label{sec:qsoimp}

\begin{figure} 
\begin{centering}
\begin{tabular}{c}
{\epsfxsize=8truecm \epsfysize=8truecm \epsfbox[35 170 550 675]{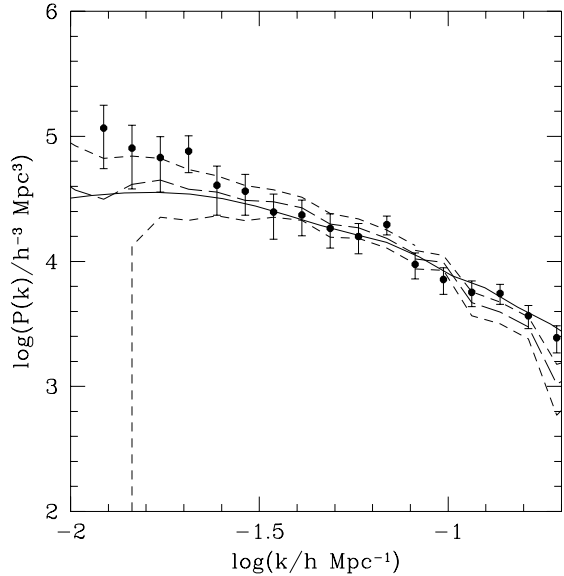}} \\
\end{tabular}
\caption {The points show the power spectrum from the
10k QSO catalogue with the 
$\Lambda$ cosmology assumed. The solid line shows the real space input
power spectra to the {\it Hubble Volume} simulation. 
The dashed lines in show the mock catalogue redshift space power
spectrum and errors from Figure \ref{fig:pkgp}(b), discussed in the
text.}
\label{fig:pkpred}
\end{centering}
\end{figure}

As discussed previously, it is the power spectrum of the mass density
field that is predicted by models of large scale structure. If the
bias between the dark matter and QSOs can be written as P(k)$_{\rm
QSO}$ = $b^2$P(k)$_{\rm mass}$, with $b$ a constant, which could be
the case on large scales (Coles 1993; Cole et al 1998; 
Mann, Peacock \& Heavens 1998),
it is meaningful to compare the shapes of mass and QSO power spectra.

However, we find that the current incompleteness may be introducing a
slight bias into the shape of the QSO power spectrum. The shape of the power
spectrum from the mock catalogues with the incompleteness imprinted on
them is slightly steeper than the input power spectrum to the
simulation. 
This is shown in Figure \ref{fig:pkpred}. The solid line shows the
input power spectrum, the long dashed line shows the power spectrum from
the mock catalogues with the current angular selection function
imprinted on them (as seen in Figure \ref{fig:pkgp}(b), the short
dashed lines are the 1$\sigma$ errors). The points show the QSO power
spectrum calculated assuming the $\Lambda$ cosmology. On the smallest
scales, the dashed line is inconsistent with the solid line
at the 1$\sigma$ level, although on these small scales the FKP errors may 
underestimate the true error. Therefore, we fit the model power
spectra to the QSO power spectrum, assuming the $\Lambda$ cosmology,
down to 60$h^{-1}$Mpc (log($k / h$Mpc$^{-1}$)=$-1.$) only. 
We have no mock catalogues with the EdS
cosmology to test if incompleteness affects the shape of the QSO power
spectrum from the 10k QSO catalogue, assuming the EdS
cosmology. However, we wish to fit the models to the data over the
same range of log($k/h {\rm Mpc}^{-1}$) so we fix the lower 
limit for the EdS case
to be log($k / h$Mpc$^{-1}$)=$-0.9$ (50$h^{-1}$Mpc).
We set the limit on large scales to
be the scale where the geometry of
the survey affects the shape of the power spectrum. This corresponds
to 300$h^{-1}$Mpc (log($k / h$Mpc$^{-1}$)=$-1.7$) 
in the EdS cosmology and 400$h^{-1}$Mpc (log($k / h$Mpc$^{-1}$)=$-1.8$)
in the $\Lambda$ cosmology.

We compare the QSO power spectra to model CDM power spectra with
different values of the shape parameter $\Gamma$, introduced in Section 
\ref{sec:gamma}. We
assume the fit to the CDM transfer function given by
Bardeen et al. (1986). The power spectrum is calculated at $z=1.4$ rather
than at $z=0$ to approximate the dark matter power spectrum averaged
over the lightcone. We assume a value of $\sigma_8$ that matches
the cluster normalisation for each cosmology given by
Eke, Cole \& Frenk (1996). For each cosmology, the model power spectra are
$\chi^2$ fitted to the QSO power spectrum to find the factor, $b_{\rm
eff}$, required to match the two as well as possible.

We choose values of $\Gamma$ in the range $0.05< \Gamma < 0.5$ which
more than covers the range of values that fit current galaxy power
spectra, such as the APM real space power spectrum 
(Eisenstein \& Zaldarriaga 2001). 
We note that $\Lambda$CDM models with $\Gamma \gsim$0.3 are not
physically motivated but we are just comparing a range of models with
different shapes to the QSO power spectrum and we do indeed find that
models with values of $\Gamma \gsim 0.3$ give a poor fit to the QSO
power spectrum with the $\Lambda$ cosmology assumed.

\begin{figure} 
\begin{centering}
\begin{tabular}{c}
{\epsfxsize=8truecm \epsfysize=8truecm \epsfbox[35 170 550 675]{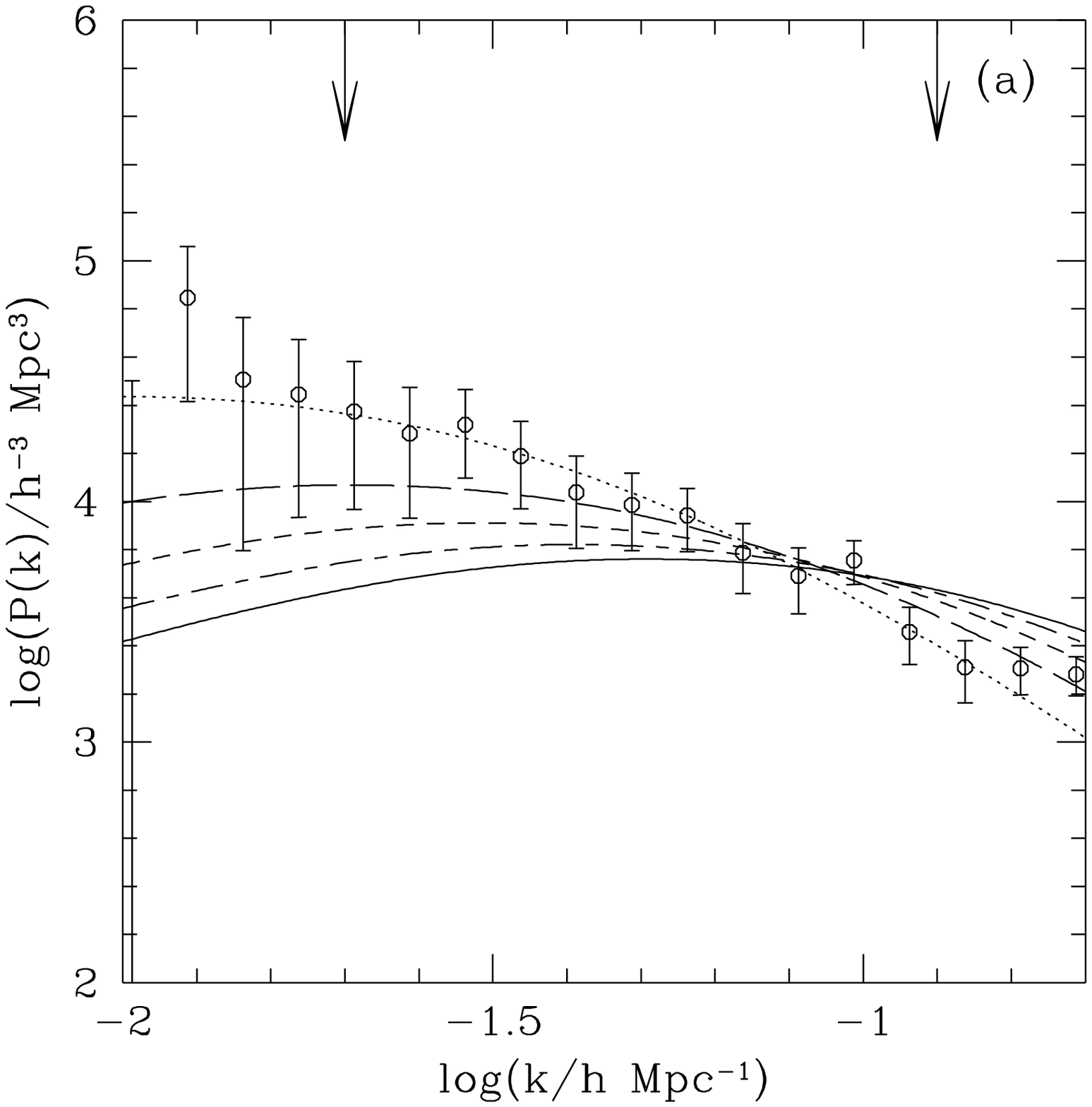}} \\
{\epsfxsize=8truecm \epsfysize=8truecm \epsfbox[35 170 550 675]{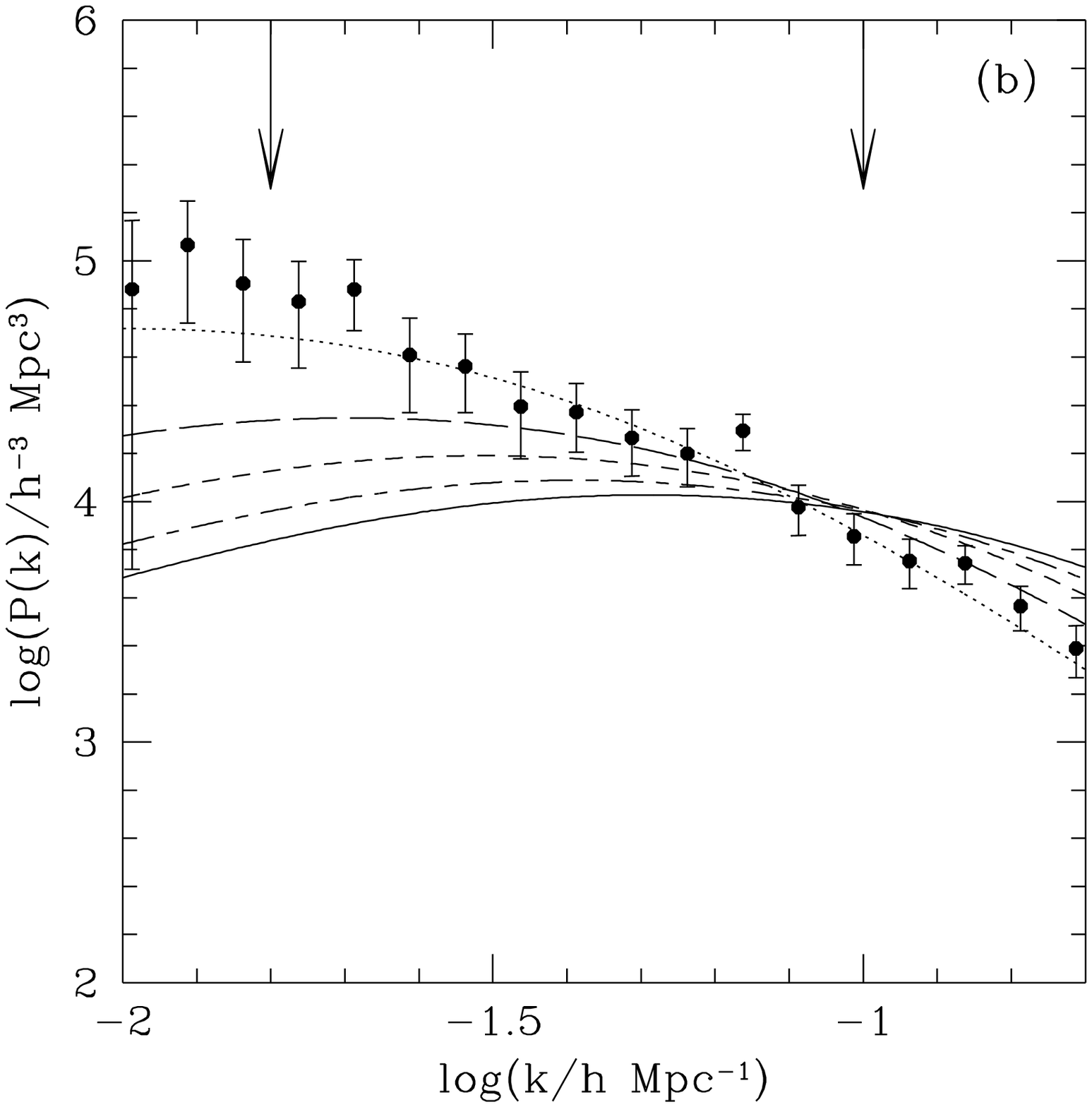}} \\
\end{tabular}
\caption{Comparison of different model power spectra to the QSO power spectra for the two choices of cosmology. The solid circles in panel (a) show the QSO P(k) with the $\Lambda$ cosmology assumed and the open circles in panel (b) show the QSO P(k) with the EdS cosmology assumed. In both panels the matching lines show the same $\Gamma$ values with $\Gamma$=0.5 (solid line), 0.4 (short - long dashed), 0.3 (short dashed), 0.2 (long dashed) and 0.1 (dotted). Each line is plotted at the amplitude that gives the best fit to the QSO power spectrum. The arrows indicate the range of scales over which the fit was made.}
\label{fig:models}
\end{centering}
\end{figure}

In Figure \ref{fig:models}, we show the QSO power spectrum with the
EdS cosmology (open circles, a) and with the $\Lambda$ cosmology
(solid circles, b). The lines have different values of $\Gamma$:
$\Gamma$=0.5 (solid), $\Gamma$=0.4 (short - long dashed), $\Gamma$=0.3
(short dashed), $\Gamma$=0.2 (long dashed) and $\Gamma$=0.1 (dotted).
Models with $\Gamma$=0.5 are ruled out at more than 3$\sigma$ assuming
either cosmology. This is in agreement with results from galaxy
surveys. More large scale power was found in the galaxy correlation
function from the APM Survey than expected from the standard CDM
$\Gamma$=0.5 model (Efstathiou, Sutherland \& Maddox 1990). 
This led to variants of the CDM
model, such as $\Lambda$CDM and $\tau$CDM to be developed.

In both cases, the best agreement between the models and the QSO power
spectra is found with a model with low values of $\Gamma$. Models with
$\Gamma$=0.12$\pm$0.10 best fit the shape of the QSO power spectrum assuming the
EdS cosmology and
models with lower values of $\Gamma = 0.11 \pm 0.1$ best fit the QSO
power spectrum assuming the $\Lambda$ cosmology. The input power 
spectrum to the $\Lambda$CDM {\it Hubble Volume} has a shape of 
$\Gamma$=0.17. This model is therefore consistent with the data at the 
1$\sigma$ level. 

The results found here are in reasonable agreement with those of 
Croom et al. (2001). They find that models with $\Gamma$=0.5 
do not provide a good fit to the correlation function measured from 
the 10k QSO catalogue assuming the EdS and the $\Lambda$
cosmology. Models with $\Gamma$=0.1-0.2 are required to 
fit the correlation function with the $\Lambda$ cosmology assumed. However, models with 
$\Gamma$=0.2-0.4 best fit the correlation function if the EdS 
cosmology is assumed.

This test is quite simplistic as possible effects of the
incompleteness have not been included in the shape of the model power
spectrum, although by fitting to a limited range of scales we should
have reduced the effect that this can have on the models that fit the
power spectrum. When 2QZ is finished, we will be able to compare the
models over a wider range of scales and produce stronger constraints
on the range of acceptable $\Gamma$ parameters.

In order to thoroughly test models of structure
formation, full N-body simulations, such as the {\it Hubble Volume}
are required. However, due to the large volume of 2QZ,
large amounts of super-computing time are required to carry out such
simulations so it is currently not possible to run them for a wide
range of cosmological models. Physically motivated models for
selecting QSOs within the simulation are also required as a linear
bias between the mass and the QSOs may be an over simplification, even
on these large scales.

\section{Discussion and Conclusions}
\label{sec:qsoconc}

We have presented the first QSO power spectrum analysis using the 10k
catalogue from the 2dF QSO Redshift Survey. The QSO power spectrum has a
power law form with P$_{\rm QSO}(k)\propto k^{-1.3 \pm0.15}$ which 
appears to extend to
large scales. More large scale power is seen in the QSO power spectra than
in the APM power galaxy spectrum even.

The clustering environment observed for radio-quiet QSOs in 
CCD imaging experiments is close to that of optically selected galaxies, 
(Ellingson, Yee \& Green 1991; Smith, Boyle \& Maddox 1995, 1999; 
Croom \& Shanks 1999)
suggesting that the biasing of QSOs may be no
stronger than it is for galaxies; 
at the r$>$50h$^{-1}$Mpc scales considered here the
assumption that the QSO-mass bias is scale independent may therefore
be reasonable. Therefore, the fact that we do not see a turnover on
scales up to 300$h^{-1}$Mpc places a strong constraint on the value of
the shape parameter, requiring $\Gamma \lsim 0.25$.  We find that the
large scale power seen in the QSO power spectrum is best fit by CDM
models with $\Gamma$ as low as $0.1 \pm 0.1$.

A power spectrum analysis of mock QSO catalogues, created from the Virgo
Consortium's Hubble Volume $\Lambda$CDM simulation, with $\Gamma=0.17$ as
input, produces a result which is statistically consistent with the data.
The analysis of the mock catalogues indicates that we are able to produce
a first robust measurement of the QSO power spectrum over a wide range of
scales and thus the above results for $\Gamma$ are unlikely to be
dominated by systematic effects due to the current incompleteness of the 2QZ
catalogue.

We note that the low $\Gamma$ values measured here from the shapes of the 
QSO power
spectra are in some disagreement with the best fit parameters as measured from
the Boomerang and MAXIMA CMB power spectra (Balbi 2000, de Bernadis 2000). 
Bond et al. (2000) measures
$\Omega_{\rm b} h^2$=0.03 and $\Omega_{\rm CDM} h^2$=0.17 
for $\Omega_{\rm m} h^2$=0.2 in a
spatially flat model. For h=0.65 this implies
$\Gamma=\Omega_{\rm m}h=0.3$ which is rejected at 2$\sigma$ in Figures 
\ref{fig:models}(a, b). 
It will be interesting to see if this discrepancy persists as more CMB and
2QZ data accumulate.

We measure the QSO power spectrum in two bins of redshift and find that
the clustering can only evolve slowly as a function of redshift as QSOs
with a median redshift of $\bar{z}$=1.0 have a clustering amplitude
consistent with that of QSOs with a median redshift of $\bar{z}$=1.8. This
is consistent with the results of Croom et al. (2001), who find little
evolution in the amplitude of QSO clustering as a function of redshift if
the EdS cosmology is assumed and only slow evolution if a $\Lambda$
cosmology is assumed.

The QSO power spectrum, which is therefore broadly independent of
redshift, lies close in amplitude to the power-spectrum of local, 
optically selected galaxies. The range of scales where the two overlap
is limited, but once the 2dF Galaxy Redshift Survey is finished, an 
interesting comparison between power spectra of optically selected
QSOs and galaxies over a wide range of scales will be possible. The
power spectrum of rich galaxy clusters measured by Tadros, 
Efstathiou \& Dalton (1998)
has a far higher clustering amplitude than the QSO power spectrum,
again suggesting that the QSO-mass bias is much closer to galaxies    
than galaxy clusters.

We tentatively detect a `spike' feature at
$\approx 90 h^{-1}$Mpc assuming the $\Lambda$ cosmology or
$\approx 65 h^{-1}$Mpc assuming the EdS cosmology. This feature
appears to reproduce in both the NGC and SGC strips and 
in independent redshift bins but
its statistical significance is still marginal. An investigation using
CMBFAST indicates that the spike is seen at a $\gsim$25 per cent smaller
wavenumber than the second peak caused by acoustic oscillation in the
pre-recombination baryon-radiation fluid. Somewhat higher values of
$\Omega_b$ than usually assumed in $\Lambda$CDM models may also be needed
to fit the amplitude of the feature. 
It will be interesting to see if this feature persists in the
final 25k 2QZ survey P(k). Since it lies well into the linear regime where it 
will evolve little with redshift under the correct assumed
cosmology it could therefore act as an important probe of world 
models.

The QSO power spectrum is currently measured to an accuracy of $\lsim
30$\% over the range of scales 60$ \lsim r \lsim 200 h^{-1}$Mpc for either
the EdS or $\Lambda$ cosmology adopted here.  On these scales, we predict
the errors will be approximately 1.5 times smaller in the completed 2QZ
catalogue.  The power spectrum from the 25k catalogue will also be
measurable out to scales of $\lsim 600 h^{-1}$Mpc. Thus, when these data
become available, there is the promise of placing even stronger
constraints on the large scale structure of the Universe and hence on the
underlying cosmological model.

\section*{Acknowledgments} 
We would like to recognise the considerable efforts of all the AAT staff responsible for the operation of the 2dF instrument which has made this survey possible. We would also thank our colleagues on the 2dF galaxy redshift survey team, in particular Gavin Dalton and Steve Maddox, for facilitating many of the survey observations and John Peacock and Shaun Cole for useful comments. We also thank the Virgo Consortium, especially Adrian Jenkins and Gus Evrard, for providing the {\it Hubble Volume} simulations. FH and NL acknowledge the receipt of a PPARC studentship.


\begin{thebibliography}{}
\bibitem[]{}
Bailey~J. \& Glazebrook~K., 1999, 2dF User Manual, Anglo Australian Observatory
\bibitem[]{}
Balbi~A. et al., 2000, ApJ, 454, 1
\bibitem[]{}
Bardeen~J.~M., Bond~J.~R., Kaiser~N. \& Szalay~A., 1986, ApJ, 304, 15
\bibitem[]{}
Baugh~C.M. \& Efstathiou~G., 1993, MNRAS, 265, 145
\bibitem[]{}
Baugh~C.M. \& Efstathiou~G., 1994, MNRAS, 270, 183
\bibitem[]{}
Bond~J.~R. et al., 2000, astro-ph/0011378
\bibitem[]{}
Boyle~B.~J., 1986, PhD Thesis, University of Durham
\bibitem[]{}
Boyle~B.~J., Shanks~T., Croom~S.~M., Smith~R.~J., Miller~L., Loaring~N. \& Heymans~C., 2000, MNRAS, 317, 1014
\bibitem[]{}
Cole~S.~M., Fisher~K.~B. \& Weinberg~D.~H., 1995, MNRAS, 267, 785
\bibitem[]{}
Cole~S.~M., Hatton~S.~J., Weinberg~D.~H. \& Frenk~C.~S., 1998, MNRAS, 300, 945
\bibitem[]{}
Coles~P., 1993, MNRAS, 262, 1065.
\bibitem[]{}
Colless~M., 1998, `Looking Deep in the Southern Sky', eds Morganti~R. \& Couch~W.~J., ESO/Australia Workshop, Springer, Pg 9
\bibitem[]{}
Crampton~D., Le F{\`e}vre~O., Lilly~S.~J. \& Hammer~F., 1995, ApJ, 455, 96
\bibitem[]{}
Croft~R.~A.~C., Weinberg~D.~H., Katz~N. \& Hernquist~L., 1998, ApJ, 495, 44
\bibitem[]{}
Croom~S.~M., 1997, PhD Thesis, University of Durham
\bibitem[]{}
Croom~S.~M. \& Shanks~T., 1996, MNRAS, 281, 893
\bibitem[]{}
Croom~S.~M. \& Shanks~T., 1999, MNRAS, 303, 411
\bibitem[]{}
Croom~S.~M., Shanks~T., Boyle~B.~J., Smith~R.~J., Miller~L. \& Loaring~N., 1998, Evolution of Large Scale Structure: From Recombination to Garching
\bibitem[]{}
Croom~S.~M., Shanks~T., Boyle~B.~J., Smith~R.~J., Miller~L., Loaring~N.~S. \& Hoyle~F., 2001, MNRAS, 325, 483  
\bibitem[]{}
de Bernardis~P. et al., 2000, Nature, 404, 955
\bibitem[]{}
Davis~M. \& Faber~S.~M., 1998, in `Wide Field Surveys in Cosmology', Editions Frontieres, ISBN 2-8 6332-241-9, 333
\bibitem[]{}
Efstathiou~G., Sutherland~W.~J. \& Maddox~S.~J., 1990, Nature, 348, 705
\bibitem[]{}
Eisenstein~D.~J. \& Zaldarriaga~M., 2001, ApJ, 546, 2
\bibitem[]{}
Eke~V.~R., Cole~S.~M. \& Frenk~C.~S., 1996, MNRAS, 282, 263
\bibitem[]{}
Ellingson~E., Yee~H.~K.~C. \& Green~R.F., 1991, ApJ, 371, 49
\bibitem[]{}
Feldman~H.~A., Kaiser~N. \& Peacock~J.~A., 1994, ApJ, 426, 23  
\bibitem[]{}
Frenk~C.S.. et al., 2000, astro-ph/0007362
\bibitem[]{}
Gunn~J.~E. \& Weinberg~D.~H., 1995, in proceedings of the 35th Herstmonceux workshop, Cambridge University Press, Cambridge
\bibitem[]{}
Hatton~S.~J., 1999, PhD Thesis, University of Durham
\bibitem[]{}
Hewitt~P.~C., Foltz~C.~B. \& Chaffee~F.~H., 1995, AJ, 109, 1498
\bibitem[]{}
Hoyle~F., 2000, PhD Thesis, University of Durham 
\bibitem[]{}
Hoyle~F., Baugh~C.~M., Shanks~T. \& Ratcliffe~A., 1999, MNRAS, 309, 659
\bibitem[]{}
Kaiser~N., 1987, MNRAS, 227, 1
\bibitem[]{}
La Franca~F., Andreani~P. \& Cristiani~S., 1998, ApJ, 497, 529
\bibitem[]{}
Le F\`{e}vre~O.  et al., 1998, in `Wide Field Surveys in Cosmology', Editions Frontieres, ISBN 2-8 6332-241-9, 333
\bibitem[]{}
Lin~H., Kirschner~P., Shectman~S.~A., Landy~S.~D., Oemler~A., Tucker~D.~L. \& Schechter~P.~L., 1996, ApJ, 471, 617
\bibitem[]{}
Mann~R.~G., Peacock~J.~A. \& Heavens~A.~F., 1998, MNRAS, 293, 209.
\bibitem[]{}
Meiksin~A. \& White~M, 1999, MNRAS, 308, 1179
\bibitem[]{}
Meiksin~A., White~M. \& Peacock~J.~A., 1999, MNRAS, 304, 851
\bibitem[]{}
Miller~L., et al., 2001, in preparation
\bibitem[]{}
Osmer~P.~S., 1981, ApJ, 247, 762
\bibitem[]{}
Peacock~J.~A., 2000, astro-ph/0002013
\bibitem[]{}
Peacock~J.~A. et al., 2001, in Proceedings of the 20th Texas Symposium on Relativistic Astrophysics
\bibitem[]{}
Schlegel~D.~J., Finkbeiner~D.~P. \& Davis~M., 1998, ApJ, 500, 525
\bibitem[]{}
Seljak~U. \& Zaldarriaga~M., 1996, ApJ, 469, 437
\bibitem[]{}
Shanks~T., Fong~R., Boyle~B.~J. \& Peterson~B.~A., 1986, In IAU Symp No 119 on Quasars, eds Swarup~G. \& Kapahi~V.~K., Reidel, Dordrecht, Holland
\bibitem[]{}
Shanks~T., Fong~R., Boyle~B.~J. \& Peterson~B.~A., 1987, MNRAS, 271, 753
\bibitem[]{}
Shaver~P.~A., 1984, A\&A, 136, L9
\bibitem[]{}
Smith~R.~J., 1998, PhD Thesis, University of Cambridge
\bibitem[]{}
Smith~R.~J., Boyle~B.~J. \& Maddox~S.~J., 1995, MNRAS, 277, 270
\bibitem[]{}
Smith~R.~J., Boyle~B.~J. \& Maddox~S.~J., 2000, MNRAS, 313, 252
\bibitem[]{}
Smith~R.~J., Croom~S.~M., Boyle~B.~J., Shanks~T., Miller~L. \& Loaring~N.~S., 2001, MNRAS submitted
\bibitem[]{}
Steidel~C.~C., Adelberger~K.~L., Givalisco~M., Dickinson~M. \& Pettini~M., 1999, ApJ, 519, 1
\bibitem[]{}
Steidel~C.~C., Pettini~M. \& Hamilton~D., 1995, ApJ, 110, 2519
\bibitem[]{}
Sugiyama~N., 1995, ApJS, 100, 281
\bibitem[]{}
Sutherland~W. et al., 1999, MNRAS, 308, 289
\bibitem[]{}
Tadros~H. \& Efstathiou~G., 1996, MNRAS, 282, 1381
\bibitem[]{}
Tadros~H., Efstathiou~G. \& Dalton~G., 1998, MNRAS, 296, 995
\bibitem[]{}
Wilson~G., Kaiser~N. \& Luppino~G.~A., 2001, ApJ, 556, 601
\bibitem[]{}
Yee~H.~K.~C., Morris~S.~L., Lin~H., Carlberg~R.~G., Hall~P.~B., Sawicki~M., Patton~D.~R., Wirth~G.~D., Ellingson~E. \& Shepherd~C.~W., 2000, ApJS, 129, 475

\end{thebibliography}
\end{document}